\documentclass[useAMS,usenatbib]{mnras}%
\usepackage[T1]{fontenc}
\usepackage[utf8]{inputenc}
\usepackage{ae,aecompl}
\usepackage{times}
\usepackage{amsmath}
\usepackage{amssymb}
\usepackage[english]{babel}
\usepackage[varg]{txfonts}
\usepackage{graphicx}
\usepackage[percent]{overpic}
\usepackage{tikz}
\usetikzlibrary{decorations.pathreplacing,external}
\usepackage{pdfpages}
\usepackage{flafter}
\usepackage{multirow}
\usetikzlibrary{arrows}
\usepackage{booktabs}
\usepackage[]{natbib}
\usepackage{color}
\usepackage{gensymb}
\usepackage[]{placeins}

\newcommand{\ud}{d}

\newcommand{\eq}[1]{Eq.~(\ref{#1})}
\newcommand{\fig}[1]{Fig.~\ref{#1}}
\newcommand{\msun}{{\ensuremath{M_\odot}\,}}

\title[]{Gravitational-wave Signals From Three-dimensional Supernova Simulations With Different Neutrino-Transport Methods}
\author[H.~Andresen et al.]{
  H.~Andresen$^{1}$\thanks{E-mail: haakon.andresen@aei.mpg.de}, R.~Glas$^{2,3}$, H.-Th.~Janka$^{2}$,
  \\ $^1$Max Planck Institute for Gravitational Physics (Albert Einstein Institute), Am M\"uhlenberg
  1, Potsdam-Golm, D-14476, Germany \\
  $^{2}$Max-Planck-Institut f\"ur Astrophysik, Karl-Schwarzschild-Str. 1, D-85748 Garching, Germany \\
 $^3$Excellence Cluster ORIGINS, Boltzmannstr.~2, D-85748 Garching, Germany }
\def\LaTeX{L\kern-.36em\raise.3ex\hbox{a}\kern-.15em T\kern-.1667em\lower.7ex\hbox{E}\kern-.125emX}

\begin{document}
\maketitle
\begin{abstract}
We compare gravitational-wave (GW) signals from eight three-dimensional simulations of core-collapse
supernovae, using two different progenitors with zero-age main sequence masses
of 9 and 20 solar masses ($M_\odot$). The collapse of each progenitor was simulated four times, at
two different grid resolutions and with two different neutrino transport methods, using the
\textsc{Aenus-Alcar} code. The main goal of this study is to assess the validity of recent concerns that the
so-called ``Ray-by-Ray+'' (RbR+) approximation is problematic in core-collapse simulations and can
adversely affect theoretical GW predictions. Therefore, signals from simulations using RbR+ are
compared to signals from corresponding simulations using a fully multidimensional (FMD) transport
scheme. The 9\msun progenitor successfully explodes, whereas the 20\msun model does not. Both the
standing accretion shock instability and hot-bubble convection develop in the postshock layer of the
non-exploding models. In the exploding models, neutrino-driven convection in the postshock flow is
established around 100 ms after core bounce and lasts until the onset of shock revival. We can,
therefore, judge the impact of the numerical resolution and neutrino transport under all conditions
typically seen in non-rotating core-collapse simulations.
We find excellent qualitative agreement
in all GW features. We find minor quantitative differences between simulations, but find no
systematic differences between simulations using different transport schemes.
Resolution-dependent differences in the hydrodynamic behaviour
of low-resolution and high-resolution models have a greater impact on the GW signals
than consequences of the different transport methods. Furthermore, increasing the resolution
decreases the discrepancies between models with different neutrino transport.
\end{abstract}

\section{Introduction}
During the last decade theoretical predictions of GWs emitted during the deaths of massive stars
based on fully three-dimensional (3D) simulations
\citep{wongwathanarat_10b,takiwaki_12,hanke_13,takiwaki_14,lentz_15,wongwathanarat_15, mueller_15b,
  melson_15a,roberts_16,mueller_17,kuroda_16b,kuroda_16,kuroda_17,summa_18,oconnor_18,
  ott_18,glas_19} have been published by several authors
\citep{mueller_e_12,yakunin_15,kuroda_16,andresen_17,oconnor_18,andresen_19,powell_19,
  radice_19,vartanyan_19,powell_20,mezzacappa_20}.  For non-rotating and slowly rotating
progenitors, two signal components have been found to dominate the GWs. Emission below 250 Hz
(low-frequency emission) has been associated \citep{kuroda_16,andresen_17} with the so-called
standing accretion shock instability (SASI) \citep{blondin_03} that develops in a subset of all
simulations (see for example \citealp{hanke_13}).  Recently, \cite{powell_20} presented signals from
a non-exploding simulation.  In one of their models the GWs generated by SASI activity reached
frequencies of $\sim 400$ Hz, due to the small average shock radius at late times seen in this
model. In addition to the low-frequency GWs, oscillations of the proto-neutron star (PNS) lead to
emission above 300 Hz \citep{marek_08,murphy_09,mueller_13}.  This signal component, which we will
refer to as high-frequency emission, is seen in all modern simulations and exhibits a secular
increase in emission frequency as time progresses.

Complementary to the theoretical efforts, the GW-astronomy community has developed methods to
specifically search for and characterise supernova GW signals in both current and future laser
interferometers
\newpage
\citep{logue_12,gossan_16,powell_16,vanputten_16,powell_17,
  gill_18,astone_18,powell_18,roma_19,suvorova_19}. The LIGO-Virgo collaboration has performed
follow-up searches for optically observed supernovae during their first observation run
\citep{ligo_sn_search} and later updated their results with data from their second observation run
\citep{ligo_sn_search2}.  The null observations of these searches put constraints on the energy
emitted in GWs by the supernovae and, as a consequence, constrains the emission models and
progenitor properties.  Observational efforts are hampered by the fact that we can not construct
template banks, which is the standard for compact binary mergers. This is due to the stochastic nature
of the fluid dynamics, the uncertainties in the underlying physics, and the high computational cost
of simulations.  To improve detection prospects of supernova GWs, the core-collapse supernova
modelling community must provide predictions that are as accurate as possible and for which
systematic effects in the signals are well understood.

The spectral properties of the individual signal components and how they depend on the evolution of
global features of the simulations are by now reasonably well understood
\citep{sotani_17,torres-forne_18,morozova_18,torres-forne_19a,torres-forne_19b,sotani_20}; the
prevalence, strength and validity of different signal components are still, however, debated in the
literature (see the discussion in \citealp{radice_19} and \citealp{mezzacappa_20}).  The roots of
this discussion are the uncertainties that remain in supernova modelling because the properties of
the GW signals are inherently tied to the underlying model dynamics.

One possible source of model differences can be the treatment of neutrino transport and the
associated neutrino-matter interactions. The energy deposited behind the shock by neutrinos
radiated away from the PNS plays a crucial role in launching the supernova shock, which initially
stalls at about one hundred kilometres from the centre.  Furthermore, the development of
hydrodynamic instabilities, the properties of the PNS, and the conditions in the postshock region
are closely connected to the details of the neutrino radiation. Radiation transport schemes used in
simulations today are approximate solutions of the full Boltzmann problem, which is currently
not feasible to solve in 3D simulations. A method that has been extensively employed is the RbR+
neutrino-transport scheme \citep{rampp_02,buras_06b,bruenn_13}, which has the advantage of being
well suited for large numerical simulations. The formulation lends itself to parallelisation and
scales well with the number of computation nodes. However, in the RbR+ approximation, only radial
neutrino fluxes are taken into account. This has been the cause of some concerns within the
core-collapse modelling community and it has been shown that the RbR+ scheme can lead to artificial
results in axis-symmetric simulations \citep{skinner_16,just_18,glas_19}. Naturally, this could, in
turn, lead to errors in the GWs predictions. However, due to the inverse cascade of energy and strict
symmetry constraints imposed in two spatial dimensions, it is not clear that results from
axis-symmetric simulations can be extrapolated to full 3D simulations. Indeed, \cite{glas_19}
compared multidimensional simulations with the RbR+ and FMD schemes and found excellent agreement in
the overall evolution of the fluid flow in the 3D simulations. The RbR+ transport scheme leads to
more pronounced local and instantaneous variations in the neutrino heating/cooling rates and
neutrino fluxes, but in 3D such regions are short lived and do not grow to
larger-scale, more coherent, flow patterns.

In this work, we will present the GW signals from the eight 3D core-collapse supernova simulations
of \cite{glas_19}.  The model set contains two different progenitors, simulated at two grid
resolutions with the RbR+ and FMD transport schemes.  Our primary goal is to investigate the
differences between GWs from simulations using the RbR+ scheme and simulations that were conducted
with FMD transport. We will perform a detailed analysis of the signals to determine if any
differences between the signals go beyond the stochastic nature of core-collapse simulations.  We
will also study how the GWs depend on the numerical resolution and compare resolution effects to
differences induced by changing the neutrino transport scheme.

In section~\ref{sec:gw} we will describe how the GWs are extracted from the simulations, summarise
current results based on 3D simulations, and show how the global properties of the model dynamics
relate to the typical features of the GW signals. The numerical methods are described in
section~\ref{sec:numerics}.  The model dynamics and GW signals of the best-resolved models are
presented in section~\ref{sec:highres}, and the discrepancies and similarities of the signals from
different simulations will also be discussed in this section. The results from the low-resolution
simulations are presented in section~\ref{sec:results_low}. We will give a summary of our main
findings and our conclusions in section~\ref{sec:disscon}.

\section{Gravitational Waves from Core-Collapse Supernovae}
\label{sec:gw}
In this section, we first give a brief description of how we extract the signals from the simulation
output.  We will then summarise current core-collapse GW theory, focusing on how the spectral
properties of the signals are connected to the global model properties. These results have been
established in the literature over the last few years and we will only give an overview together
with the appropriate references.
\subsection{Extracting the Gravitational Waves}
\label{sec:gw_ext}
The GW signals are extracted from the hydrodynamic simulations by post-processing the output data
using the quadrupole formula.  In the transverse-traceless (TT) gauge the GW tensor
can be expressed in terms of two independent components, $h_{+}$ and $h_{\times}$.  Far away from
the source, at a distance $D$, and in the slow-motion limit, the two components can be expressed in
the spherical coordinate system of the simulations as follows:
\begin{align}\label{eqT:hp}
  h_{+} = \frac{G}{c^4 D} & \Big[ \ddot{Q}_{11} (\cos^2{\phi} - \sin^2{\phi} \cos^2{\theta})
    \\ \nonumber & + \ddot{Q}_{22} (\sin^2{\phi} - \cos^2{\phi} \cos^2{\theta}) - \ddot{Q}_{33}
    \sin^2{\theta} \\ \nonumber & - \ddot{Q}_{12} (1 + \cos^2{\theta}) + \ddot{Q}_{13} \sin{\phi}
    \sin{2\theta} \\ \nonumber & + \ddot{Q}_{23} \cos{\phi} \sin{2\theta} \Big]
\end{align}
and

\begin{align}
\label{eqT:hc}
  h_{\times} = \frac{G}{c^4 D} & \Big[ (\ddot{Q}_{11} - \ddot{Q}_{22})
    \sin{2\phi}\cos{\theta}\\ \nonumber & +\ddot{Q}_{12} \cos{\theta} \cos{2\phi} - \ddot{Q}_{13}
    \cos{\phi} \sin{\theta} \Big].
\end{align}
Here $\theta$ represents the polar angle, and $\phi$ is the azimuthal angle of the spherical
coordinate system defined by the simulation grid.  We use $c$ for the speed of light and $G$ is
Newton's constant.  $\ddot{Q}_{ij}$ denotes the second-order time derivatives of the Cartesian
components of the quadrupole moment (with $i,j = 1,2,3$), and is given by
\begin{equation}
\label{eq:STFQ}
\ddot{Q}_{ij} = \mathrm{STF} \left [2 \int \ud^3 x \, \rho \left ( v_i v_j - x_i \partial_j \Phi
  \right) \right].
\end{equation}
In this form, the time derivatives of the original definition have been eliminated to avoid
numerical problems associated with second-order derivatives \citep{oohara_97, finn_89, blanchet_90}.
In \eq{eq:STFQ}, $v_i$ are the Cartesian velocity components, $x_i$ the Cartesian coordinates, the
gravitational potential is represented by $\Phi$ (including post-Newtonian corrections used in the
simulations), and $\rho$ is the local fluid density.  $\mathrm{STF}$ denotes the symmetric
trace-free projection operator.  In the following, we do not give the GW strain $h_+$ and
$h_{\times}$, but the \emph{GW amplitudes:}
\begin{align}
\label{eqT:zhchx}
A_{+} \equiv D h_{+}, \qquad A_{\times} &\equiv D h_{\times}.
\end{align}
For a more detailed derivation of the above equations see \cite{mueller_e_12} and
\cite{andresen_17}.

We compute spectrograms by applying short-time Fourier transforms (STFT) to $A_{+}$ and $A_{\times}$
individually, and we square the output before adding them together. Before plotting, we normalise
the STFT and take the logarithm. The normalisation is such that the logarithmic value lies in the range
of $(-\infty, 0]$. The same factor is applied for each model so that the plots can be directly
  compared.  In other words, we define and plot
\begin{equation} \label{eq:spectrogram}
\Pi(A_{+},A_{\times}) \equiv \log_{10} \gamma \big[ \mathrm{STFT}(A_{+})^2 +
  \mathrm{STFT}(A_{\times})^2\big],
\end{equation}
where $\gamma$ is the normalisation factor. We use $\gamma = 1/0.05$.  For a discrete signal, the
STFT is obtained by applying a discrete Fourier transform (DFT) to the signal with a sliding window.
We use the following DFT definition:
\begin{equation} \label{eq:DFT}
\widetilde{X}_k (f_k) = \frac{1}{M} \sum^M_{m=1} x_m e^{-2\pi i k m/M}.
\end{equation}
Here, $x_m$ is the discrete time-domain signal, consisting of $M$ samples. $f_k = k/T$ represents
the frequency of bin $k$, where $T$ is the signal length. The spectrograms shown in this work use a
window length of 60 ms. The signal segments are convolved with a Blackman window before the DFT is
applied, and we filter out anything above 1200 Hz or below 25 Hz in the DFTs.

The total energy, $E$, radiated in GWs by a source is given by
\begin{align} \label{eq:energy}
E &=\frac{G}{5 c^5} \int \mathrm{d}t \, \dddot{Q}_{ij} \dddot{Q}_{ij}.
\end{align}
Here $\dddot{Q}_{ij}$ are the third-order time derivatives of the quadrupole moment components.
The corresponding spectral energy density of the GWs, for a discrete time signal with duration $T$, is
\begin{equation} \label{eq:spectralenergy}
\Bigg[\frac{\Delta E}{\Delta f}\Bigg]_k =\frac{2G}{5 c^5} (2\pi f_k)^2
\Big[\big|\widetilde{\ddot{Q}_{ij}} \widetilde{\ddot{Q}_{ij}}|\Big]_k \, T^2.
\end{equation}

{The simulations we study in this work were terminated at different times, and the GW
 emission had not subsided when the simulations ended. The dependency on the signal length in
 \eq{eq:spectralenergy} is, therefore, disadvantageous for our comparison of the different GW
 signals. We circumvent this problem by calculating the spectral energy density for a time window
 determined by the duration of the shortest signal of each model set. For the s9 models, we calculate the spectral
 energy density of the GWs emitted between 50 and 420 ms post bounce, and for the s20 models
 the time window stretches from 50 to 570 ms post bounce.}

\subsection{The Characteristics of Gravitational Waves from Three-dimensional Simulations} \label{sec:gw_chr}
By now, the typical GWs emitted during the core collapse of slowly rotating (and non-rotating)
progenitors have been well established in the literature
\citep{marek_08,murphy_09,mueller_e_12,mueller_13,yakunin_15,
 kuroda_16,sotani_17,andresen_17,torres-forne_18,morozova_18,
 oconnor_18,andresen_19,powell_19,radice_19,
 torres-forne_19a,torres-forne_19b,vartanyan_19,sotani_20,mezzacappa_20}. The signal predictions
are diverse and several unique signal components have been discovered, but two of these are most
prevalent in recent GW predictions. Emission with an almost linear increase in frequency as a function of time is found above 300 Hz.
It was first seen in
two-dimensional simulations and was suggested to be associated with buoyancy processes in the
surface of the PNS \citep{marek_08,murphy_09,mueller_13}. It was shown that the central frequency of
this emission component closely traces the Brunt-V\"ais\"ala frequency in the PNS surface region
\citep{murphy_09,mueller_13}, which was also found to be the case for signals from recent
3D simulations \citep{andresen_17,morozova_18,andresen_19,oconnor_18}.

The Brunt-V\"ais\"ala frequency ($N$) is given in terms of the sound speed ($c_s$), the pressure
($P$), the gravitational potential ($\Phi$), and the density ($\rho$) as follows:
\begin{equation} \label{eq:BV}
N^2 = \frac{1}{\rho} \frac{\partial \Phi}{\partial r} \left [ \frac{1}{c_s^2} \frac{\partial
  P}{\partial r} - \frac{\partial \rho}{\partial r} \right ].
\end{equation}
Here $r$ denotes the radial coordinate of a spherical coordinate system. The quantities in
\eq{eq:BV} are not constant over the outer PNS layer, the radial variance of these quantities as
well as local perturbations cause the GW emission to spread out around the average value of $N$.
While spatially averaging \eq{eq:BV} over the PNS surface layer results in a good match between
the average $N$-value and the peak-frequency measured from the spectrograms, it is not straight forward to directly
discern the PNS properties based on \eq{eq:BV}. \citet{mueller_13} derived an
approximation for $N$ which facilitates a better understanding of the connection between the global
PNS properties and the typical frequency ($f_{\mathrm{GW}}^{\mathrm{b}}$, here ``b'' denotes
buoyancy) of the GWs emitted by buoyancy processes. According to \citet{mueller_13},
\begin{equation} \label{eq:fp}
 f_{\mathrm{GW}}^{\mathrm{b}} =\frac{1}{2\pi} \frac{G M_{\mathrm{PNS}}}{R_{\mathrm{PNS}}^2}
 \sqrt{1.1\frac{m_n}{\langle \varepsilon_{\bar{\nu}} \rangle }}\bigg[1-\frac{G M_{\mathrm{PNS}}}{R_{\mathrm{PNS}}
   c^2} \bigg]^2.
\end{equation}
This formulation is not only informative with regard to extracting the physical parameters of the
system from the signal, but it also provides some insights relevant for the purpose of this work.
$M_{\mathrm{PNS}}$ and $R_{\mathrm{PNS}}$ in \eq{eq:fp} represent the radius and mass of the PNS.
The average anti-electron neutrino energy is denoted by $\langle\varepsilon_{\bar{\nu}}\rangle$, which is simply
a proxy for the temperature of the PNS, and $m_{n}$ is the neutron mass. If the global properties
of the PNS do not depend strongly on the neutrino transport scheme, then there will be no reason to
expect a large difference in the average properties of the GW emission above 300 Hz.

The second signal component commonly found is associated with the SASI. Strong emission centered
around $\sim\,75$-$100$ Hz is found in simulations where the SASI dominates the postshock
flow \citep{kuroda_16,andresen_17,andresen_19,oconnor_18}. The SASI arises when perturbations
created at the shock are advected with the fluid down to the PNS. These perturbations are converted
to sound waves at the PNS surface which propagate back up to the shock and give rise to new
perturbations. Under the right conditions this cycle grows in amplitude and results in large-scale
quasi-periodic oscillations of the shock
\citep{blondin_03,blondin_06,foglizzo_07,ohnishi_06,ohnishi_08,scheck_08,guilet_12,foglizzo_15}.
The typical time scale, $\tau$, of the SASI is given by the time it takes a perturbation to be
advected from the shock-front plus the time needed for the sound waves to propagate back upstream
to the shock. Assuming spherical symmetry, we can estimate $\tau$ as follows:
\begin{equation} \label{eq:tsasi}
 \tau \approx \Bigg[ \int_{R_{\text{PNS}}}^{R_{\text{S}}}\ud r \bigg( \frac{1}{c_s} +
  \frac{1}{|v_r|}\bigg) \Bigg].
\end{equation}
In the above expression $R_{\text{S}}$ represents the average shock radius, and $v_r$ is the
advection velocity at a distance $r$ from the origin. We can neglect the term $1/c_s $ since the
sound speed in the postshock region is much larger than the advection velocity, which to first order
is a linear function of the radial position, $v_r = v_{\text{PS}} r / R_{\text{S}}$. Here
$v_{\text{PS}} = -\beta^{-1} \sqrt{G M_{\text{PNS}}/R_{\text{S}}}$ is the postshock velocity
\citep{mueller_14}. The dimensionless parameter $\beta$ is the ratio of the postshock and preshock
density. The inverse of the typical SASI time scale, with a factor of two accounting for frequency
doubling, sets the characteristic frequency ($f_{\mathrm{SASI}}^{\mathrm{GW}}$) of GWs associated
with SASI activity. After solving the integral and inverting the expression, one finds
\begin{equation} \label{eq:fsasi}
 f_{\mathrm{SASI}}^{\mathrm{GW}} \approx 2 \, \beta^{-1} \sqrt{\frac{G
   M_{\mathrm{PNS}}}{R_{\mathrm{S}}^3}} \bigg[ \ln
  \big(R_{\mathrm{S}}/R_{\mathrm{PNS}}\big)\bigg]^{-1}.
\end{equation}
For a more detailed discussion of the SASI and the typical time scale on which the instability
develops and operates we refer the reader to \citet{foglizzo_07}, \citet{scheck_08},
\citet{mueller_14}, and \citet{janka_17}. Again we see that the typical frequency depends on the
properties of the PNS, but also the average shock radius. Typically, the shock
trajectory exhibits a larger degree of variation between different simulations than the PNS
does. Assuming that the PNS radius is 30 km and that the average shock radius is 100 km, a 10 per
cent increase in the average shock radius would correspond to a $\sim 24$ per cent decrease in
$f_{\mathrm{SASI}}^{\mathrm{GW}}$. A 5 per cent larger shock radius results in roughly a 10 per
cent lower emission frequency. Since the low-frequency emission is spread out in a roughly 200 Hz
broad frequency range, differences in the trajectory of the shock caused by changing the
neutrino-transport scheme must be systematic and sustained over long periods to have a noticeable effect of the GW
signal.
\section{Numerical Methods} \label{sec:numerics}
The core-collapse simulations which this paper is based on were presented in \citet{glas_19}, they
were performed with the \textsc{Aenus-Alcar} code \citep{obergaulinger_phd,just_15,just_18}. The
module \textsc{Aenus} \citep{obergaulinger_phd,just_15,just_18} uses a Godunov-type method to solve
the hydrodynamical equations. The solver is a directionally unsplit finite-volume scheme in
spherical coordinates ($r$, $\theta$, $\phi$). The high-density equation of state used was SFHo
\citep{steiner_13}, which was extended to be applicable for temperatures down to ${10^{-3}
 \ \mathrm{MeV}}$ \citep{glas_19}. \textsc{Aenus} describes self-gravity in terms of a
one-dimensional potential, which includes relativistic corrections to the Newtonian potential (case
A of \citealp{marek_06}).

\textsc{Alcar} is a fully multidimensional neutrino-radiation solver that employs a two-moment
scheme to evolve the two lowest angular moments of the Boltzmann equation \citep{just_18}. Three
neutrino species (electron neutrinos, electron antineutrinos, and a third species representing the
remaining neutrino flavours), in 15 energy bins logarithmically spaced up to $400 \ \mathrm{MeV}$,
were evolved in the simulations of \cite{glas_19}. The RbR+ simulations were performed by setting
non-radial flux terms to zero for the duration of the simulations.

The simulations were performed with two different grid resolutions, which allowed the authors to
judge the impact of numerical resolution. The two resolutions used were
$(n_r,\ n_{\theta},\ n_{\phi}) = (320, \ 40, \ 80)$ and $(n_r, \ n_{\theta}, \ n_{\phi}) = (640,
\ 80, \ 160)$, here $n_r$, $n_{\theta}$, and $n_{\phi}$ indicate the number of grid points in the
radial, polar, and azimuthal directions, respectively. We will refer to the former set-up as the
low-resolution simulations and use the label ``L'' (for low) and the latter set-up will be labelled
with ``H'' (for high) and will be referred to as the high-resolution simulations. Note that the
grids were not uniform: at the poles the last two step-sizes in the $\theta$-direction were
increased. In the high-resolution case, the step-size first increases to 4 degrees and then to 10
degrees, while for the low-resolution case the increase is first to 6 degrees and then to 12
degrees. The rest of the computational domain was covered by equally sized cells in $\theta$.

The inner ten kilometres of the grid were evolved in spherical symmetry. This simplification is
bound to have impacted the convective region within the PNS and the GWs produced there. However,
since the extent of the spherical core was kept the same in every simulation and since PNS
convection is not our primary focus, the relatively large spherical core should not affect the
conclusion of this work.
\cite{glas_19} simulated the core collapse and the post-bounce phase of two stellar progenitors.
The more massive of the two, a star with a zero-age main-sequence mass of 20 $\msun$ and with solar metallicity, fails
to explode in the simulated period of evolution. This progenitor is described in
\cite{woosley_07}. The simulations based on the second star, which is a solar metallicity star with
a zero-age main-sequence mass of 9 $\msun$, results in successful explosions. The progenitor is a
modified version \citep{sukhbold_16} of one of the progenitors described in \citet{woosley_15}.

\section{High-resolution Simulations} \label{sec:highres}
Since the high-resolution simulations are the most interesting for this work, we will focus on them
here and return to the low-resolution simulations towards the end of this
paper. Separating the resolution aspect avoids confusing the model description and the following
discussion, the reader can find a detailed and in-depth description of all the simulations in
\citet{glas_19}. In the following sections, we adopt a naming convention where a progenitor name (s9
or s20) followed by either ``-RbR'' or ``-FMD'' indicates the model simulated with the RbR+ or FMD
neutrino transport scheme, respectively. We remind the reader that a -H is appended to the model
names of the high-resolution models and that a model name ending in -L indicates a low-resolution
model.

Our discussion of the model dynamics will focus mainly on the aspects directly relevant to the GW
signals.
\subsection{Model Dynamics} \label{sec:moddym}
Changing the neutrino treatment in the high-resolution simulations does not lead to large
differences in the model dynamics. The average shock radius, the PNS radius, the neutrino heating
rate, and the prevalence of hydrodynamic instabilities are remarkably similar between the different
realisations of the post-bounce phase of both s9 and s20. However, spatial and temporal variations in
the details do occur.
\begin{figure*}
\centering \includegraphics[width=0.99\textwidth]{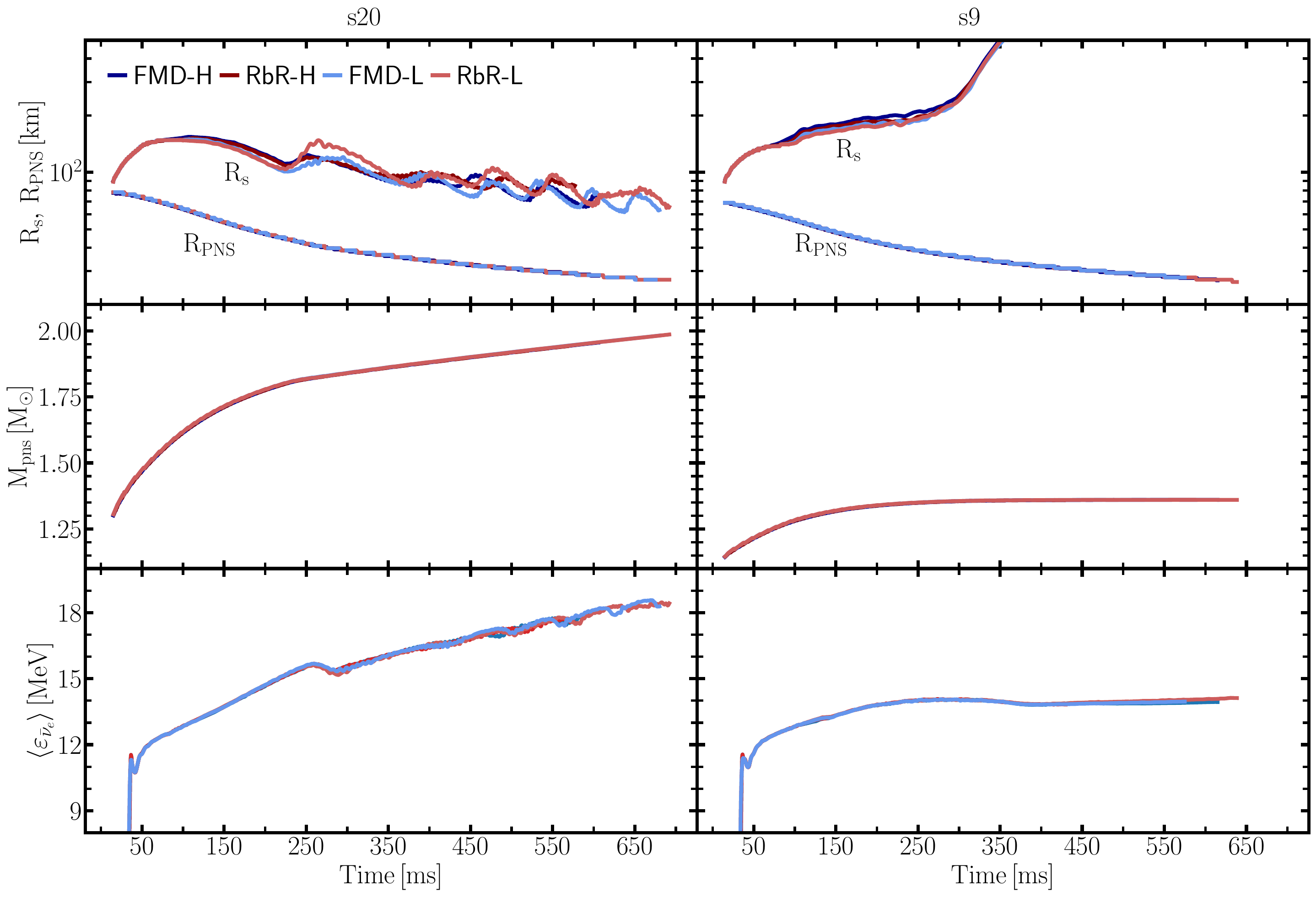}
\caption{Average shock radius ($\mathrm{R_{s}}$), PNS radius ($\mathrm{R_{PNS}}$), PNS mass
 ($\mathrm{M_{PNS}}$), and the average electron antineutrino energy ($\langle
 \varepsilon_{\bar{\nu}_e} \rangle$) as a function of time. The average shock radius and the PNS
 radius are shown in the top panels, the middle panels show the PNS mass, and the bottom panels show the average electron antineutrino energy
 extracted at a radius of 400 km in the co-moving frame.
 The left column represents the s20 models and the right column the s9 models. Time is given in ms after core bounce.}
\label{fig:pns_prop_h}
\end{figure*}
In both high-resolution simulations of the s20 progenitor, the shock front follows roughly the same trajectory, see
the top left panel of \fig{fig:pns_prop_h}. The initial expansion and subsequent stagnation of the
shock front are followed by a period of shock contraction, starting at around 100 ms post bounce and
lasting until the silicon/oxygen interface reaches the shock. As this interface passes through the
shock, approximately 200 ms after core bounce, a rapid decrease in the density ahead of the shock
leads to a second period of shock expansion which lasts roughly 50 ms. From around 250 ms after
bounce, the shock gradually contracts until the end of the simulation. There are transient periods
of expansion, but they are immediately followed by shock contraction. The conditions in the
postshock layer are best described as a non-linear combination of SASI and neutrino-driven
convection; the two instabilities operate concurrently with varying strength. The SASI tends to push
the shock further out, which create more favourable conditions for convection. The shock starts to retreat
as the SASI gives way to convection. The contraction of the shock reduces the advection time scale, which
leads to the reappearance of the SASI. These processes repeat cyclically and
oscillations in the average shock radius depend on the details of these highly complicated cycles.
Alternating periods of shock expansion and contraction are seen in both s20-FMD-H and s20-RbR-H, but
the cycles do not entirely coincide in the two models.

Models s9-RbR-H and s9-FMD-H result in successful supernova explosions. Shock revival starts
$\sim$300 ms after core bounce, see the right top panel of \fig{fig:pns_prop_h}. Before this time,
the average shock radius steadily increases until the conditions for runaway shock expansion are
met. The s9 progenitor is characterised by a region of low density immediately around the degenerate
core, which leads to low accretion rates in the two simulations. This favours the growth of
convective activity and disfavors the SASI, due to the resulting large advection time scales. While
strong SASI does not develop in models s9-RbR-H and s9-FMD-H, \citet{glas_19} reported that large-scale
convective plumes in the postshock layer lead to dipole deformation of the shock front. These
deformations can first be seen around 100 ms after bounce, and they reach their peak between 250 and
300 ms post bounce, see the bottom panel of Fig.~13 in \citet{glas_19}. After the onset of runaway
shock expansion, the accretion rate onto the PNS is further reduced. As in the case of the s20
models, the global properties of models s9-FMD-H and s9-RbR-H agree very well.

The properties of the PNS are virtually unaffected when changing the neutrino-transport scheme, and
this is true for both progenitors. The radius and mass of the PNS, which are important for
determining the properties of GWs, are unchanged between the FMD and RbR+ runs, see
\fig{fig:pns_prop_h}. Additionally, the luminosities and average energies of the emitted neutrinos
show little variation between the runs with different neutrino-transport schemes.
From around 400 ms after bounce, the luminosities and energies of the electron neutrinos and
antineutrinos undergo variations of $\sim \, 10$-$20\%$. The bottom panels of \fig{fig:pns_prop_h}
show the average energies of electron antineutrinos extracted at a radius of 400 km, for all four
high-resolution simulations. We do not show luminosity plots since they are not directly relevant
for the discussion about GW characteristics, they are shown in Fig.~7 and Fig.~14 of
\citet{glas_19}. The temporal variations in the average properties of the neutrinos, seen in models
s20-FMD-H and s20-RbR-H, are caused by the alternating periods of SASI and convection in the
postshock layer and the resulting variations in the conditions near the PNS surface, see
\citet{glas_19}.
\subsection{Gravitational Waves from the High-resolution Simulations}
\label{sec:results}
\begin{figure*}
\centering \includegraphics[width=0.99\textwidth]{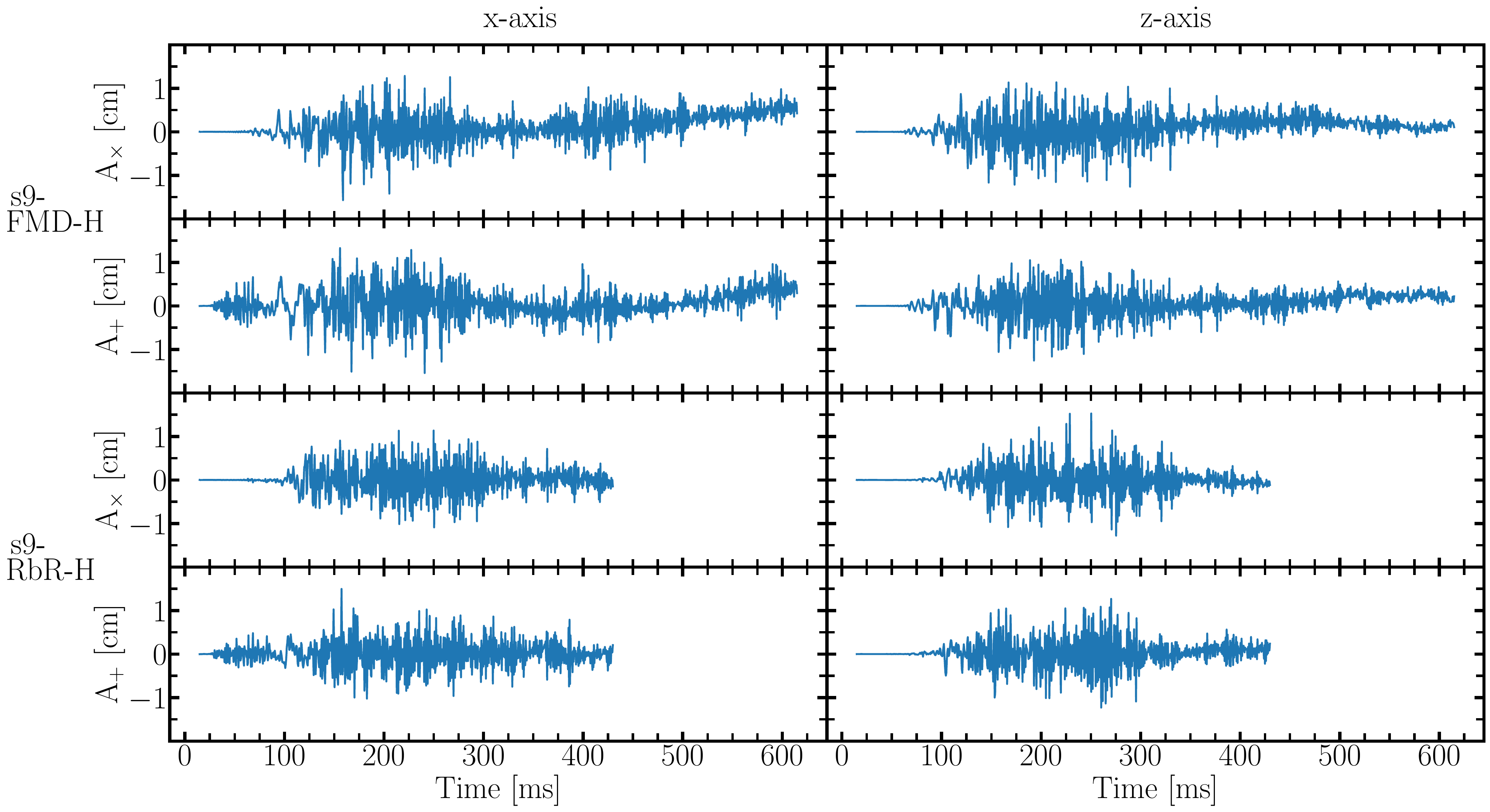}
\caption{GW amplitudes $A_+$ and $A_\times$ as functions of time after core bounce for models
  s9-FMD-H (top two rows) and s9-RbR-H (bottom two rows). The two columns show the amplitudes for
 two observers, one along the z-axis (right) and one along the x-axis (left).}
\label{fig:s9amps}
\end{figure*}

\begin{figure*}
\centering \includegraphics[width=0.99\textwidth]{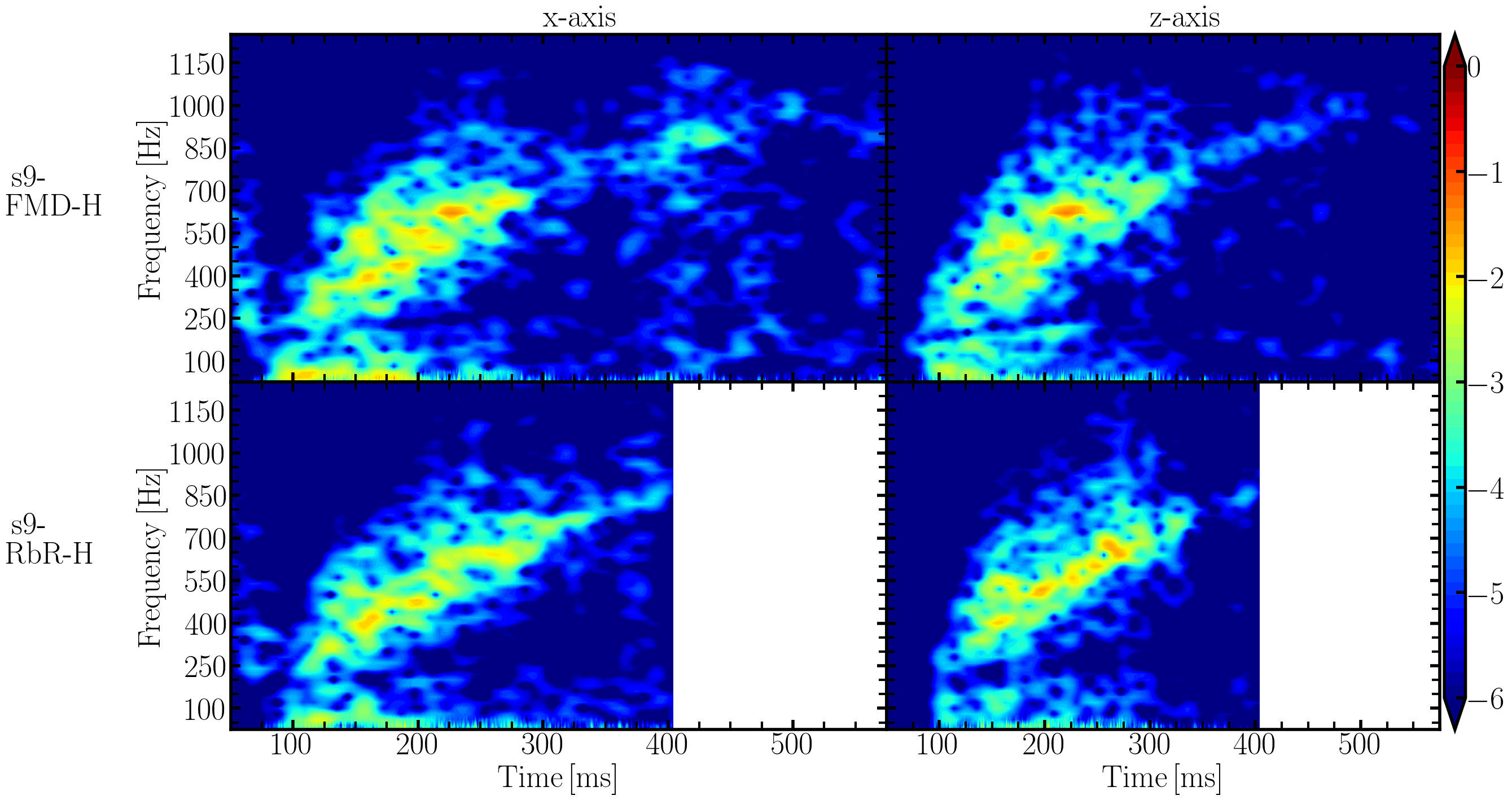}
\caption{Spectrograms, as defined by \eq{eq:spectrogram}, for models s9-FMD-H (top row) and s9-RbR-H
 (bottom row). Time is given in ms after core bounce. The two columns show the amplitudes for
 two observers, one along the z-axis (right) and one along the x-axis (left). The colour scale
 is logarithmic. The rate at which data were saved to disk in the simulations means that we have a Nyquist frequency of 1000
Hz, which results in some aliasing (see the last paragraph of section \ref{sec:disscon}). }
\label{fig:s9specs}
\end{figure*}

\begin{figure*}
\centering \includegraphics[width=0.99\textwidth]{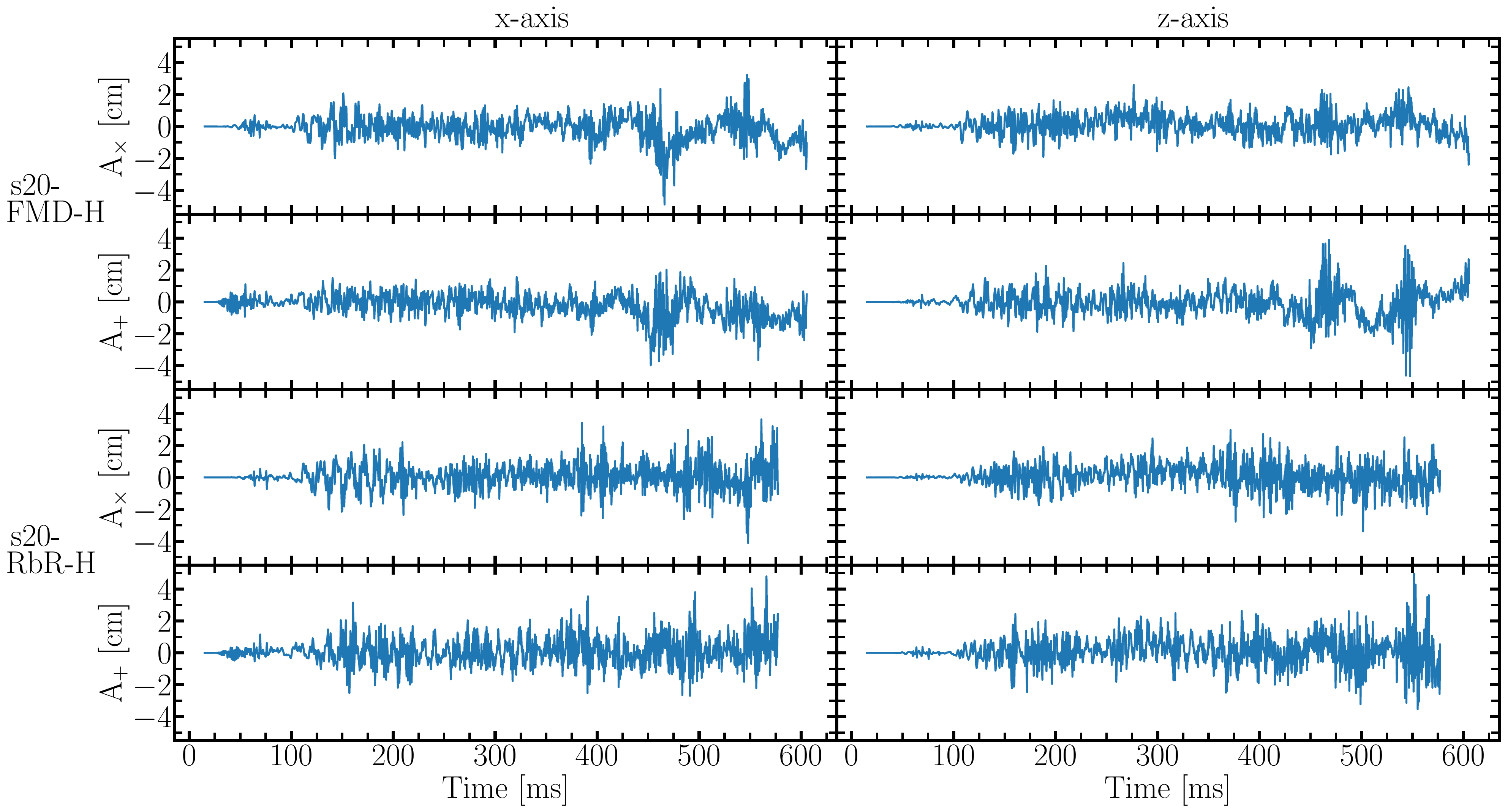}
\caption{GW amplitudes $A_+$ and $A_\times$ as functions of time after core bounce for models
 s20-FMD-H (top two rows) and s20-RbR-H (bottom two rows). The two columns show the amplitudes for
 two observers, one along the z-axis (right) and one along the x-axis (left).}
\label{fig:s20amps}
\end{figure*}

\begin{figure*}
\centering \includegraphics[width=0.99\textwidth]{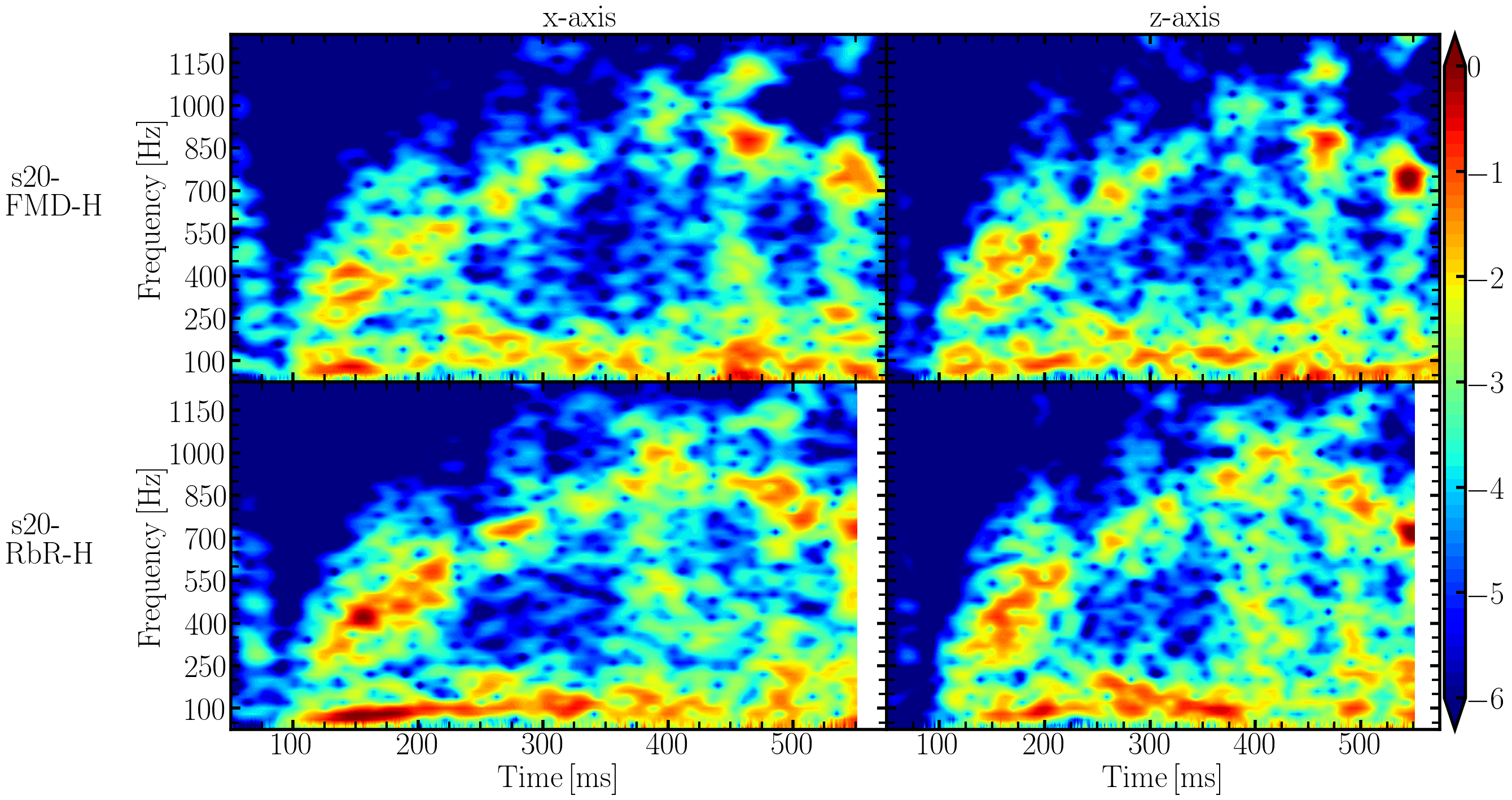}
\caption{Spectrograms, as defined by \eq{eq:spectrogram}, for models s20-FMD-H (top row) and
 s20-RbR-H (bottom row). Time is given in ms after core bounce. The two columns show the amplitudes for
 two observers, one along the z-axis (right) and one along the x-axis (left). The
 colour scale is logarithmic. The rate at which data were saved to disk in the simulations means that we have a Nyquist frequency of 1000
Hz, which results in some aliasing (see the last paragraph of section \ref{sec:disscon}).}
\label{fig:s20specs}
\end{figure*}

The GW amplitudes and the corresponding spectrograms, for two
different observer orientations, from the high-resolution s9 models
are shown in \fig{fig:s9amps} and \fig{fig:s9specs}, respectively. The signals and the spectrograms
from the high-resolution s20 models are shown
in \fig{fig:s20amps} and \fig{fig:s20specs}, respectively. 
The signals are in general very similar to what has recently been reported in the literature (see
for example \cite{kuroda_16,andresen_17,andresen_19,oconnor_18,powell_19,powell_20,mezzacappa_20}).
In all the models, the typical GW amplitudes are on the order of a few centimetres, see
\fig{fig:s9amps} and \fig{fig:s20amps}. Since vigorous SASI activity does not develop in the
simulations of the less massive progenitor, the signals from s9-RbR-H and s9-FMD-H consist mainly of
emission above 300 Hz (\fig{fig:s9specs}). However, s9-RbR-H and s9-FMD-H weakly emit low-frequency
GWs.
The emission in the s9 models continues after shock revival, but the amplitudes
are reduced by roughly a factor of two. This reduction of GW emission supports the findings of
\citet{radice_19}, who found that GWs are mainly excited in the PNS surface region by turbulent
downflows from the postshock layer. \cite{andresen_17} and \cite{andresen_19}, on the other hand,
reported that overshooting of convective plumes into the PNS surface layer from the convectively
unstable layer below was the primary source of GW emission from the PNS.

Violent SASI activity develops in models s20-FMD-H and s20-RbR-H, which is reflected in the GW
signals by emission below 250 Hz (\fig{fig:s20specs}). Since the postshock flow alternates between
being SASI and convectively dominated, the emission associated with SASI activity is intermittent
and relatively weak when, for example, compared to model m15fr of \citet{andresen_19}. The spotty
emission below 250 Hz is clearly visible in all the panels of \fig{fig:s20specs}. Model s20-RbR-H has a
period of sustained and strong low-frequency emission between 100 and 200 ms after core bounce
(bottom left panel of \fig{fig:s20specs}) which is weaker in model s20-FMD-H. The reduction of this
emission in the FMD model is partly due to the viewing angles we have chosen. A similar, but still
somewhat weaker, signal component emerges in model s20-FMD-H when changing the viewing angle.
\begin{figure*}
\centering \includegraphics[width=0.99\textwidth]{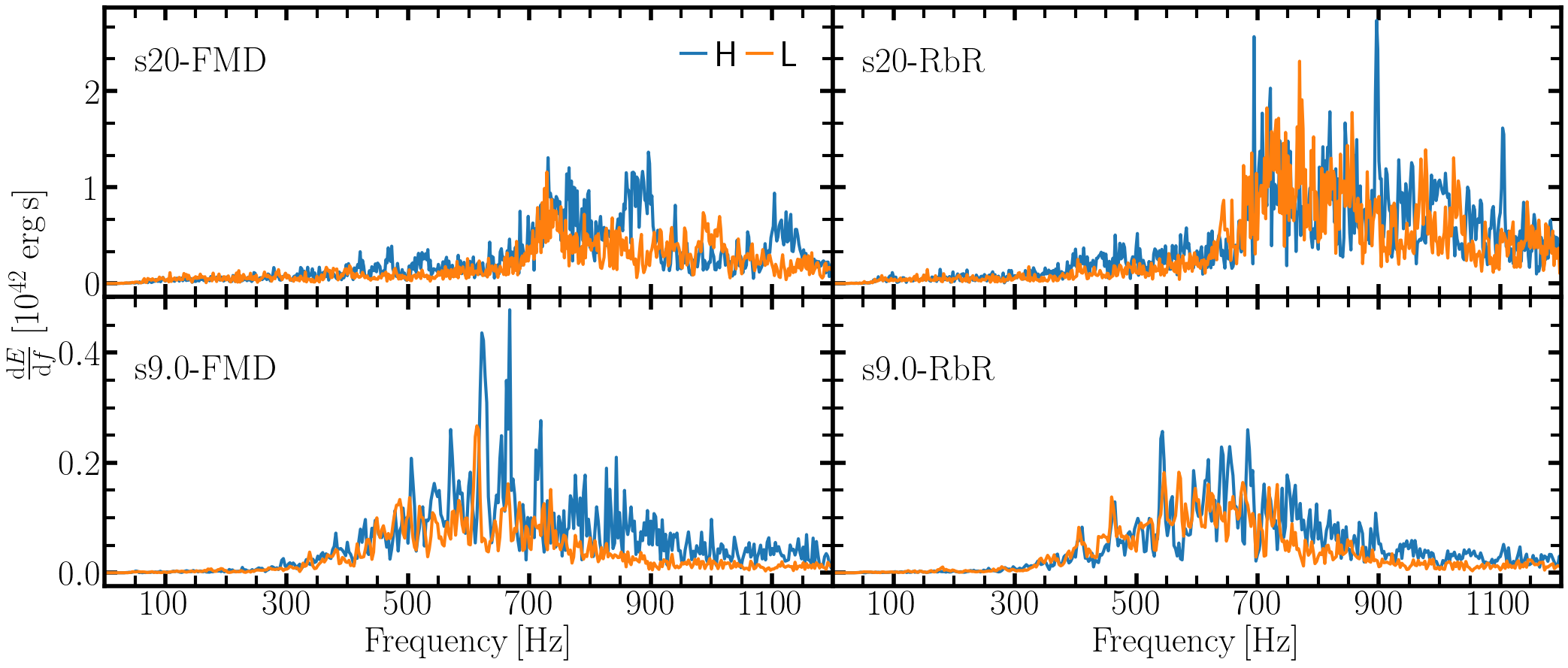}
\caption{{Total time-integrated spectral energy density (\eq{eq:spectralenergy}) of the GWs
    for all eight models. The top row shows results for the s20 simulations, which were obtained from the emission between 50 and 570 ms post bounce.
    The results of the s9 simulations are shown in the bottom row, calculated from the GWs emitted between 50 and 420 ms post bounce.
    The right column shows results from simulations with the RbR+ scheme. The left column shows the results from simulations with the FMD scheme.
    Blue lines represent the best-resolved models
  (s20-FMD-H, s20-RbR-H, s9-FMD-H, and s9-RbR-H) and the orange represent lines the low-resolution models
  (s20-FMD-L, s20-RbR-L, s9-FMD-L, and s9-RbR-L).}}
\label{fig:spectrum}
\end{figure*}

\begin{figure*}
\centering \includegraphics[width=0.99\textwidth]{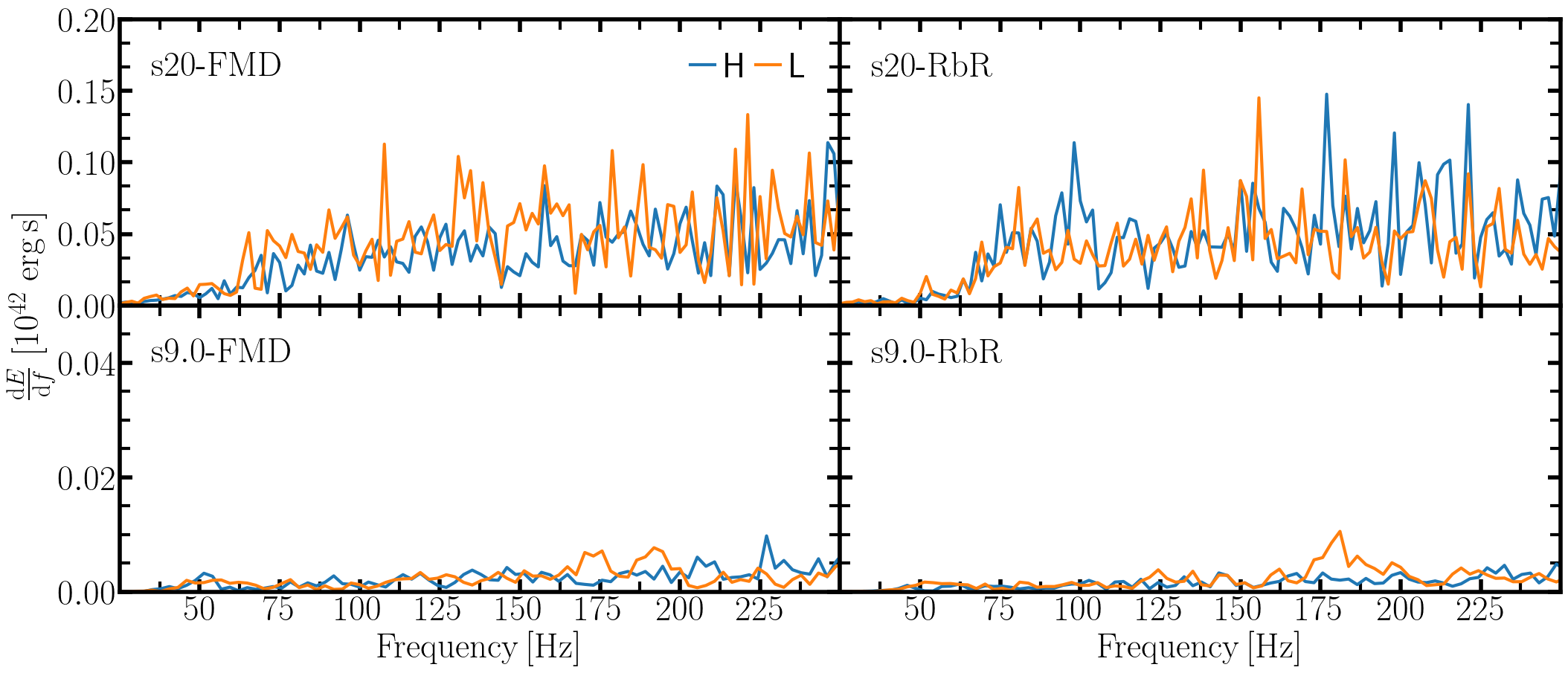}
\caption{Same as \fig{fig:spectrum}, but the with the abscissa adjusted to highlight
    the details of the low-frequency emission.}
\label{fig:spectrum_low}
\end{figure*}

\begin{figure}
\centering \includegraphics[width=0.49\textwidth]{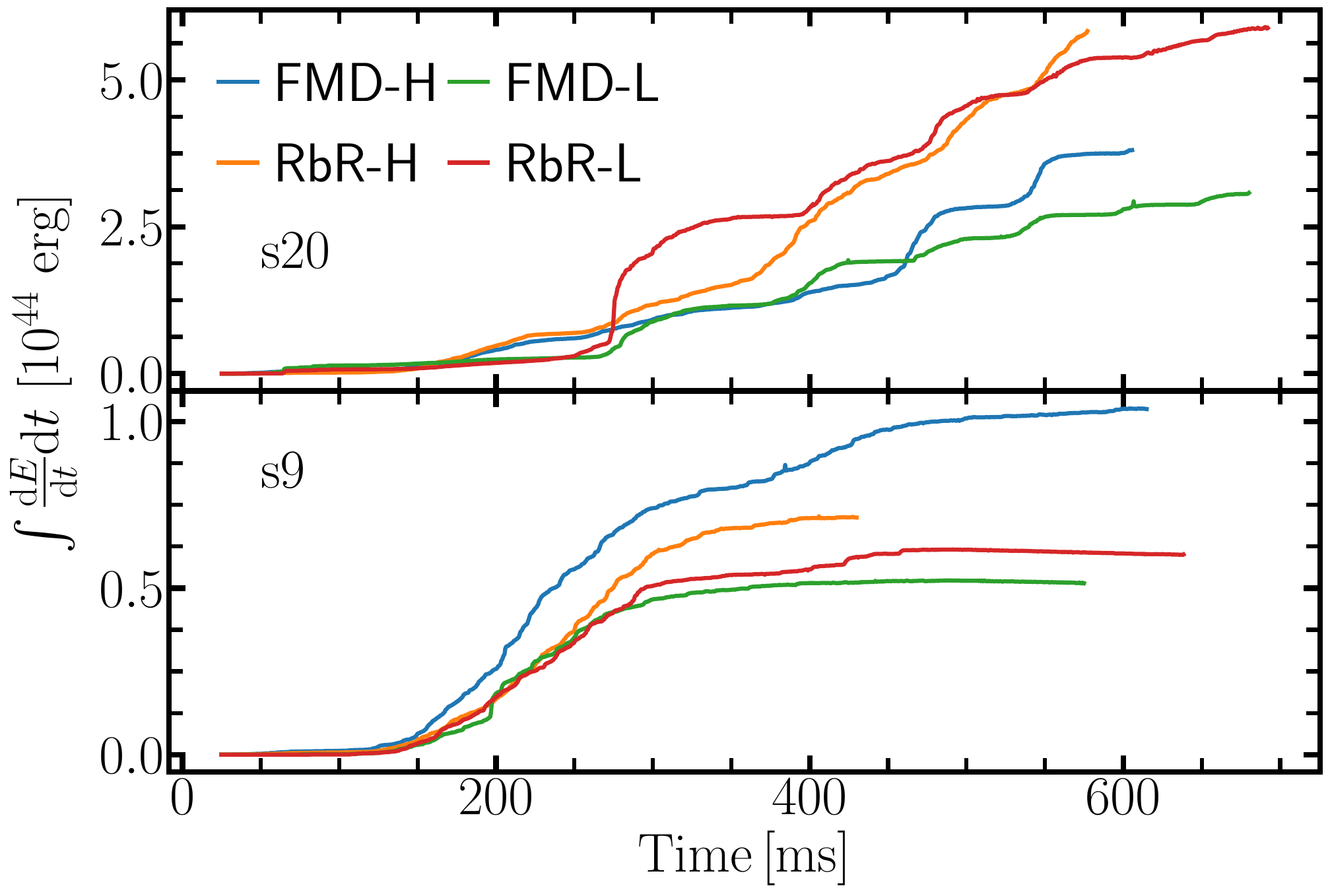}
\caption{Total time-integrated energy emitted in GWs (\eq{eq:energy}) as a function of time for all
 eight models. The top panel shows results for the s20 simulations. The results of the s9
 simulations are shown in the bottom panel. The x-axis gives the time after core-bounce. The
 curves show the emitted energy from the beginning of the simulations until the time indicated on
 the x-axis. }
\label{fig:energyt}
\end{figure}

\begin{figure*}
\centering \includegraphics[width=0.99\textwidth]{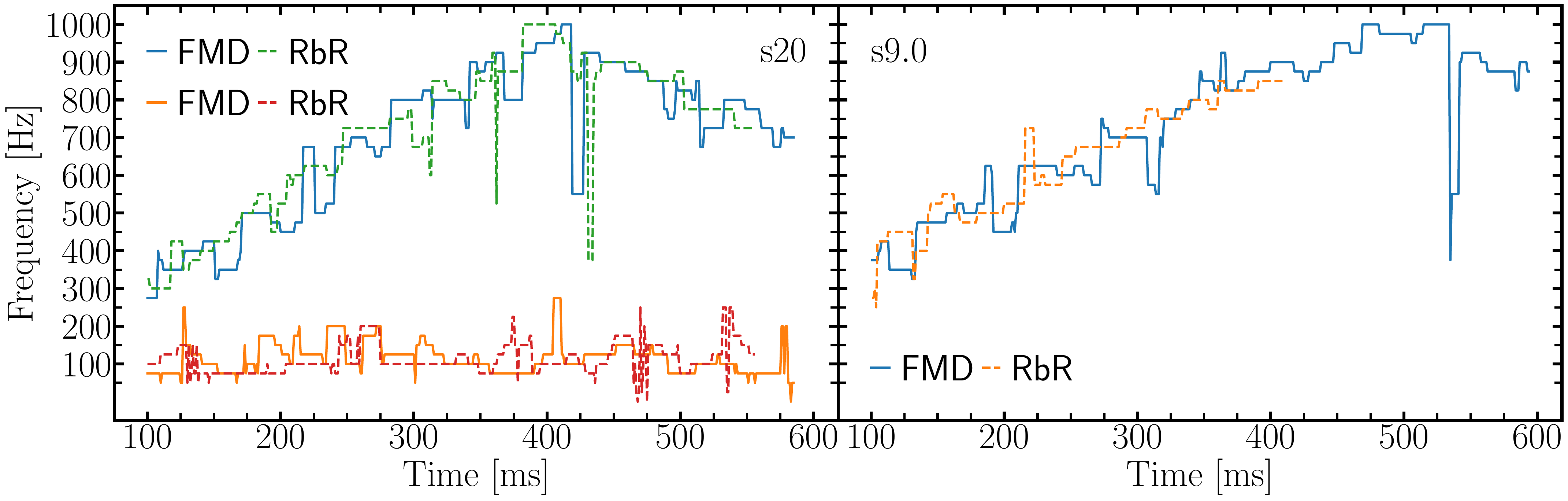}
\caption{The peak-frequency of the two main signal components as a function of time after core bounce.
  The left panel represents the models s20-RbR-H and s20-FMD-H. The right panel represents
  s9-RbR-H and s9-FMD-H. Dashed lines indicate simulations using the RbR+ scheme. Solid lines represent simulations using FMD transport.
  In the left-hand side panel, we show four lines, the orange and red curves represent the central frequency of the emission below 250 Hz and the blue and
  green curves represent the peak-frequency of the emission above 250 Hz. In the right panel, we only show two lines,
  since the s9 models do not have strong low-frequency GW emission.
  The peak frequencies are extracted from the Fourier analysis of the signals.}
\label{fig:FMT-RbR}
\end{figure*}

In \fig{fig:spectrum} and \fig{fig:spectrum_low} we show the time-integrated spectral energy density of the GW emission.
The time integration was done over intervals determined by the duration of the shortest simulation
for each of the investigated progenitors, see section~\ref{sec:gw_ext}. \fig{fig:spectrum} shows the spectral energy density
for a broad frequency range, while \fig{fig:spectrum_low}
shows the spectral energy density at frequencies relevant for the low-frequency emission.
The emission from model s20-FMD-H is dominated by three broad peaks, at $\sim$ 750 Hz, 850 Hz,
and 1150 Hz. Inspecting the spectrograms (\fig{fig:s20specs}) reveals that the peaks at 850 Hz and
1150 Hz correspond to the same physical event, a burst of emission can be seen at approximately
450 ms post bounce. The emission at 850 Hz should be emitted at 1150 Hz, but aliasing causes part of the emission to
be reflected down to lower frequencies. The emission peak at 750 Hz can also be traced back to one single emission period,
taking place around 550 ms post bounce. The emission of model
s20-RbR-H is spread out across a range of frequencies from around 300-1000 Hz. Several narrow peaks with amplitudes $\sim\,2$
times greater than the surrounding emission pierce the relative flat spectrum of the RbR+ model (\fig{fig:spectrum}).
These peaks represent short bursts of emission at fixed frequencies and
such spikes have been associated with individual downflows impinging on the PNS surface
\citep{marek_08,murphy_09,mueller_13}. We do not find direct counterparts to each of these peaks in the spectrograms.
However, the spectrograms are only shown for two distinct observer directions and it is not surprising that not every observer can see every emission
feature. The larger spectral energy density of model s20-RbR-H, compared to s20-FMD-H, is reflected in the total emitted energy (\fig{fig:energyt}).
Model s20-RbR-H emitted about 30 per cent more energy during the simulated time than s20-FMD-H did in the same period.
At intermediate frequencies, between the two main emission bands, we witness more emission in model
s20-RbR-H than in s20-FMD-H, which is also visible in \fig{fig:s20specs}.

The strong early-time emission in model s20-RbR-H is visible in top right panel of
\fig{fig:spectrum_low} as a broad and prominent peak centered around 100 Hz. The FMD high-resolution model also has a peak around
100 Hz, but it is not as broad and pronounced as in the RbR+ model. The energy spectrum of s20-FMD-H is
flatter than the spectrum of model s20-RbR-H. In the latter we can clearly see enhanced power centered around 200 Hz.
Comparing the upper and lower panels of \fig{fig:s20specs}, one can see slightly more emission above 100 Hz between
250 and 400 ms post bounce in model s20-RbR-H than from s20-FMD-H. It is important to keep in mind that
the energy spectra are weighted by the square of the frequency and it is not unexpected that particular
features are more (or less) visible in the spectra compared to the spectrograms. 
The bottom row of \fig{fig:spectrum_low} confirms what the spectrograms show, namely that
the s9 models emit much less energy at frequencies below 300 Hz than their s20 counterparts.

The GW emission of s9-FMD-H is $\sim$ 25 per cent more energetic than the GW
emission from s9-RbR-H. In the bottom panels of \fig{fig:spectrum}, we see that model s9-FMD-H
undergoes several short bursts of GW emission and that the FMD model emits more GWs at frequencies above
900 Hz than model s9-RbR-H. Furthermore, large peaks and narrow peaks can be seen in model s9-FMD-H,
which are not present in s9-RbR-H.

The central frequencies of both emission components are remarkably similar between runs with
different neutrino treatments. In \fig{fig:FMT-RbR} we plot the peak frequencies of the two emission
components as functions of time, which were extracted from the spectrograms by finding the maxima of
the Fourier transforms in any given time window. The results for s20, shown in the left panel,
include two lines for each model. The lines around 100 Hz represent the signal associated with SASI
activity and the lines in the upper half of the panel represent the emission coming from PNS
oscillations. In the right panel, there are no lines around 100 Hz, since the SASI does not develop
in models s9-FMD-H and s9-RbR-H. We see that the emission above 300 Hz is essentially independent
of the neutrino-transport scheme, the initial value and temporal evolution of the curves agree very
well. The central frequency of the emission associated with the SASI also agrees well in models
s20-RbR-H and s20-FMD-H.

\subsection{Differences Induced by the Ray-by-Ray+ Transport Scheme }
\label{sec:difftran}
It is difficult to disentangle the fluid motions responsible for individual parts of the GW emission
from the rest of the flow and trace them throughout the simulation. Consequently, we can not
explain all the small variations in the GW emission of the different models. In some cases, it is
not straightforward to establish whether or not differences in the signals are caused
by stochastic variations or real differences induced by changing the neutrino-transport scheme. As a
consequence, some of the explanations offered in this section are likely scenarios instead of proven
theories. When more simulations with a wider range of resolutions and with different input physics
become available, some revisions might be in order.

{There is no evidence that the choice of neutrino transport scheme leads to large systematic differences in the GW emission.
  The signals from simulations with different neutrino transport have similar amplitudes. Furthermore, we found no consistent
  trends in the spectral properties of the signals in correlation with the applied neutrino transport scheme.
 The RbR simulations of the s20 progenitor are more energetic and emit more GWs at higher frequencies than their
 FMD counterparts, but for the s9 progenitor the situation is reversed.
 Model s20-RbR-H emits more GWs between the two main emission bands than model s20-FMD-H, but the difference
 is small and emission at intermittent frequencies can not be seen in the s9 models.
 The typical emission frequencies agree well in all models and this can easily be understood
 from \eq{eq:fp} and \eq{eq:fsasi} in conjuncture with \fig{fig:pns_prop_h}. The global properties of the simulations that set the typical
 emission frequencies, the average shock radius, and the mass,
 temperature (or the average electron antineutrino energy), and radius of the PNS do not significantly differ between the RbR+ and FMD models.
 We do, therefore, not see substantial
 differences in the typical properties of the signals from the models with different neutrino transport, at least not systematic ones
 that occur in all cases as a clear consequence of the employed transport method.}

\begin{figure}
\centering \includegraphics[width=0.49\textwidth]{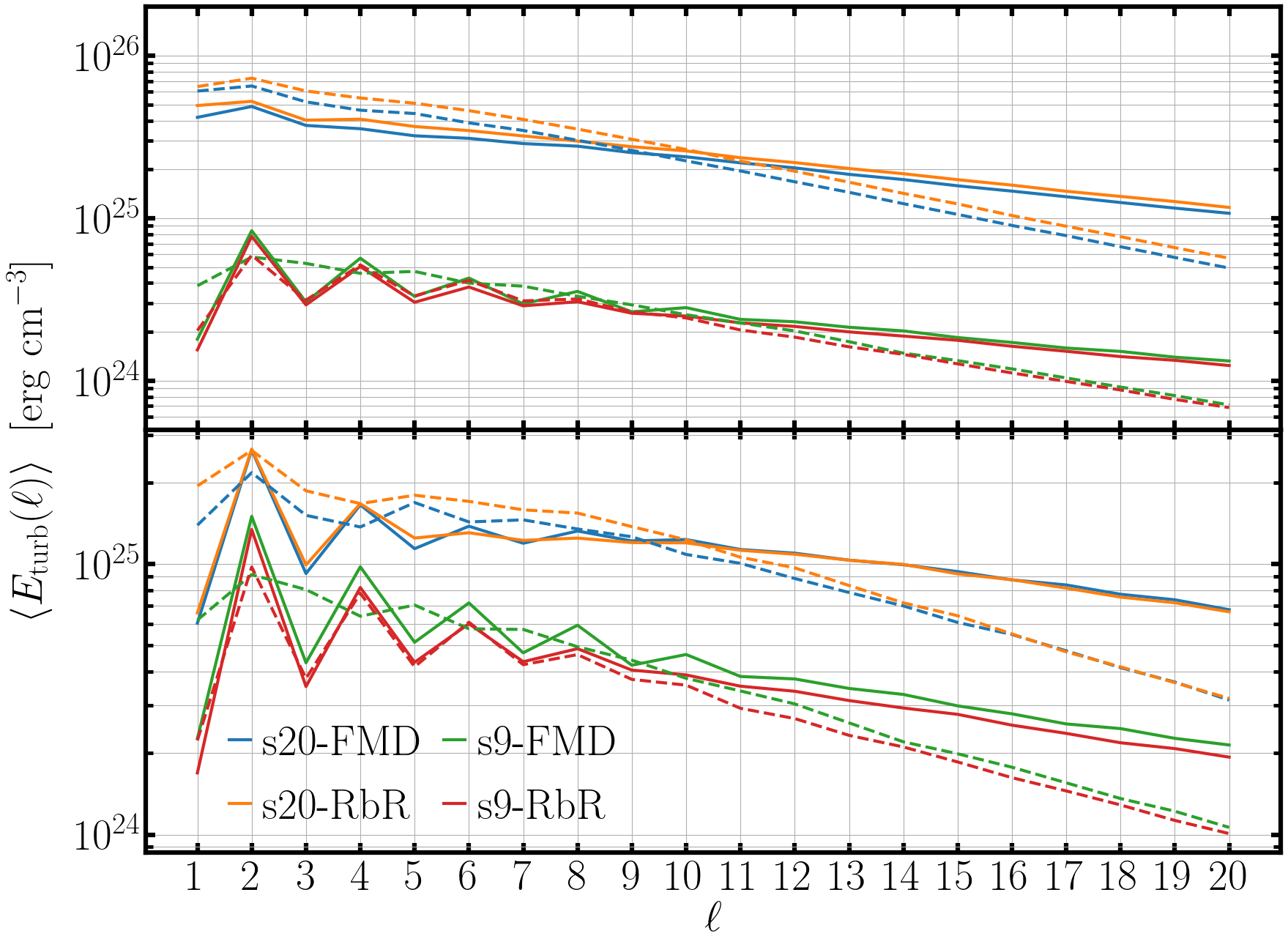}
\caption{Time averaged turbulent kinetic energy spectrum of the mass motions in the postshock layer,
  as defined by \eq{eq:ekturbavr}. The top panel shows results obtained by averaging between 50 and 420 ms post
  bounce for the s9 models and between 50 to 570 ms post bounce for the
 s20 models. The bottom panel shows the average turbulent kinetic energy spectrum in the time interval
 50 and 200 ms post bounce for all models. Solid lines denote high-resolution models and dashed lines represent the low-resolution models.}
\label{fig:energy_spectrum}
\end{figure}

The strong low-frequency emission that an observer situated in the equatorial plane
(of the simulation grid) would observe between 100 and 250 ms post bounce in model s20-RbR-H (see
\fig{fig:s20specs}) raises several questions. It is not apparent why the emission is so strong,
why it appears in a very narrow frequency range, and why it is not seen with equal strength in model
s20-FMD-H (though stronger than visible in \fig{fig:s20specs} for other observer directions).
Strong and coherent emission at low-frequencies, similar to what we see in models s20-FMD-H and
s20-RbR-H, has been connected to vigorous SASI activity in the postshock layer
\citep{kuroda_16,andresen_17,andresen_19,oconnor_18,powell_19,powell_20}.
Strong SASI activity is often associated with sizable and coherent dipolar deformation of the
stalled shock front, see for example \cite{hanke_13}. In both s20-FMD-H and s20-RbR-H,
the dipole deformation of the shock front is constrained to $\sim\,1$ per cent of the average shock radius
during the first 200 ms post bounce (see Fig. 6 of \cite{glas_19}). The fluid flow appears to be convectively dominated in both models,
as suggested by the $\chi > 3$ parameter criterion, which, however, is not rigorously accurate in a quantitative manner.
However, a quadrupole pattern is present in the flow of both s20-FMD-H and s20-RbR-H, which is evident when considering the
energy spectrum of non-radial turbulent mass motions
($E_{\mathrm{turb}}(\ell)$) in the postshock layer.
We define $E_{\mathrm{turb}}(\ell)$ as follows:
 \begin{equation} \label{eq:ekturb}
  E_{\mathrm{turb}}(\ell) = \frac{1}{V}\int^{R_{\mathrm{s}}}_{R_{\mathrm{PNS}}} r^2 \mathrm{d}r\sum_{m=-\ell}^{\ell}\Bigg|\int \mathrm{d}\Omega
  Y_\ell^m \sqrt{\rho \big(v_{\phi}^2 + v_{\theta}^2 \big)}\Bigg|^2,
 \end{equation}
 where $v_\phi$ is the fluid velocity in the azimuthal direction, $v_\theta$ is the fluid velocity in the
 polar direction, and $V$ is the total volume of the postshock layer 
 \begin{equation}
  V = \frac{4}{3}\pi (R_{\mathrm{s}}^3 - R_{\mathrm{PNS}}^3).
 \end{equation}
 Note that we define the integration limits in the radial direction by the average shock radius
 and the PNS surface (defined to be where the angle-averaged density drops below $10^{11}$ g/cm$^3$).
 In \eq{eq:ekturb}, $Y_{\ell}^{m}$ represents the spherical harmonic functions expressed in their
 real form, 
 \begin{equation} \label{eq:ylm}
  Y_{\ell}^{m} = 
  \begin{cases}
   & N_\ell^{|m|} P_{\ell}^{|m|} (\cos{\theta}) \sin{(|m|\phi)} \quad \ \mathrm{ if } \ m < 0, \\ 
   & N_\ell^0 P_{\ell}^{0} (\cos{\theta}) \quad \quad \quad \quad \quad \quad \mathrm{ if } \ m = 0, \\ 
   & N_\ell^m P_{\ell}^{m} (\cos{\theta}) \cos{(m\phi)} \quad \quad \mathrm{ if } \ m > 0.
  \end{cases}
 \end{equation}
 Here $\ell$ and $m$ are the degree and order, respectively, of the spherical harmonic functions and
 $N_\ell^m$ is given by
 \begin{equation} \label{eq:nlm}
  N_\ell^m = (-1)^m \sqrt{2} \sqrt{\frac{2\ell +1}{4\pi}\frac{(\ell - |m|)!}{(\ell + |m|)!}},
 \end{equation}
 and $P_{\ell}^m$ is the associated Legendre polynomial of degree $\ell$ and order $m$.
 The stochastic nature of turbulence causes temporal fluctuations in the turbulent energy spectrum
 and we, therefore, show the time-averaged spectrum,
 \begin{equation} \label{eq:ekturbavr}
  \langle E_{\mathrm{turb}}(\ell) \rangle = \frac{1}{t_1 - t_0}\int_{t_{0}}^{t_{1}} E_{\mathrm{turb}}(\ell) \mathrm{d}t,
 \end{equation}
 in \fig{fig:energy_spectrum}. In \eq{eq:ekturbavr}, $t_0$ and $t_1$ represent the start and the end, respectively,
 of the time window over which the turbulent energy spectrum is averaged.
 The top panel of \fig{fig:energy_spectrum} shows the turbulent energy spectrum averaged over the
 same time window used to produce the spectral energy density plots of \fig{fig:spectrum}
 (50 to 420 ms post bounce for the s9 models and 50 to 570 ms post bounce for the s20 models), and the bottom
 panel of \fig{fig:energy_spectrum} represents the average turbulent energy spectrum between 50 and 200 ms post bounce.

 For the s20 models, the large peaks at $\ell = 2$ seen in the bottom panel of \fig{fig:energy_spectrum} indicate that a 
quadrupolar SASI mode \citep{blondin_06} is superimposed on the convective flow of these models during the first $\sim$ 200 ms of their postbounce evolution.
 In models s20-FMD-H and s20-RbR-H, the high accretion at early times combined with the quadrupolar asymmetries induced by 
the SASI create ideal conditions for strong GW emission between 50 and 200 ms after core bounce.
 The early-time emission of models s20-FMD-H and s20-RbR-H is similar to, although weaker than, the low-frequency emission seen in 
model m15fr of \cite{andresen_19}. While the dynamics of m15fr and the s20 models differ substantially,
 the turbulent energy spectrum of model m15fr peaks at $\ell = 2$ \citep{summa_18}. Model m15fr of \cite{andresen_19} provides an example of a model 
where SASI develops soon after core bounce when the accretion rate is high with a turbulent energy spectrum that peaks at $\ell = 2$, and 
with strong low-frequency GW emission in a narrow frequency range.
 
Between 200 and 250 ms post bounce an $\ell = 1$ SASI mode starts to grow in the s20 models, and we start to see the typical dipolar
deformation of the shock front 
around this time. The emergence of a dipolar SASI mode around 250 ms is also
evident from the turbulent energy spectrum. Compared to the turbulent energy spectrum of the
flow between 50 and 200 ms post bounce,
the energy at $\ell = 1$ increases by a factor of 10 when averaging the turbulent energy between
50 and 570 ms post bounce, compare the top and bottom panels of \fig{fig:energy_spectrum}.

In the end, we can not conclusively answer all of the questions raised by the strong low-frequency emission at early times in the s20 models.
It seems likely that the SASI is responsible for this emission, but it is difficult to definitively prove the existence of a low-amplitude SASI mode.
Furthermore, it is unclear where the differences between the s20-RbR-H and s20-FMD-H come from.
The distribution of energy over various scales and the total energy contained in turbulent mass motions in model s20-FMD-H and s20-RbR-H are in good 
agreement during the first 200 ms after bounce (bottom panel of \fig{fig:energy_spectrum}).
There is no evidence to indicate that the differences between s20-FMD-H and s20-RbR-H, before 250 ms post bounce,
are due to anything else than stochastic variations in the fluid flow. 

Interestingly, the turbulent energy spectra of the s9 models also peak at $\ell = 2$, but the peak is
roughly a factor two smaller than the peak in the s20 models. The $\ell = 2$ peak in the turbulent energy spectra may
correspond to low-amplitude SASI activity, which never develops into large shock oscillations.
The growth of the SASI is disfavored in the s9 models due to their continuous shock expansion and corresponding
long advection time scales. However, it is more likely that the spectral peak at $\ell = 2$ is connected to large convective plumes
deforming the shock front, because the $\chi$ parameter in the s9 models is huge ( $>10$; \cite{glas_19}).

\section{Low-resolution Simulations}
\label{sec:results_low}
We now turn our attention to the four simulations carried out with an angular and radial grid
resolution which was two times coarser than in the high-resolution runs. {
\subsection{Model Dynamics}
Models s9-RbR-L and s9-FMD-L closely resemble their high-resolution counterparts. The PNS properties
do not change when reducing the resolution, and the properties of the PNS are identical in
all models of the low-resolution set (see \fig{fig:pns_prop_h}). The postshock flow of both models is dominated by neutrino-driven convection
that develops around 100 ms post bounce and lasts until the onset of rapid shock expansion around
200 ms later. Between 100 and 275 ms post bounce, the average shock radius is smaller by $\sim
1-10$ km in the low-resolution models than in models s9-RbR-H and s9-FMD-H.

The dynamics of models s20-FMD-L and s20-RbR-L are in good agreement with the high-resolution
simulations, but there are some differences in the details of the flow. The shock, both in s20-FMD-L and
s20-RbR-L, undergoes the same initial expansion as seen in the high-resolution models, peaking around
100 ms after bounce. During the expansion phase, the average shock radii of all models agree
well with each other. Once the shock starts to recede, models s20-FMD-L and s20-RbR-L exhibit lower average
shock radii than the corresponding high-resolution simulations. The smaller average shock radii in the low-resolution
models foster the SASI and this is reflected in the turbulent energy spectra of the models. Compared to the
high-resolution models, more of the turbulent energy is concentrated at $\ell = 1$ in the low-resolution models
(see \fig{fig:energy_spectrum}). The increase of energy at $\ell = 1$ is fully compatible with the higher numerical
viscosity associated with lowering the resolution. The
shock, in both models, reaches a local minimum at around 200 ms post bounce. From this point on,
model s20-FMD-L evolves similarly to the high-resolution models. The shock in model s20-FMD-L
undergoes periods of expansion and contraction, which are caused by alternating periods of SASI and
convection in the postshock layer. The same is true for model s20-RbR-L, but both the average shock
radius and its oscillations are on average larger than in the other three simulations. Furthermore,
model s20-RbR-L exhibits a phase where the average shock radius is significantly larger than in the
other three models. Between 250 and 350 ms after bounce the average shock radius of model s20-RbR-L
is $\sim $ 10-30 km larger than the shock radius of models s20-FMD-L, s20-RbR-H, and s20-FMD-H}.

\subsection{Properties of the Gravitational Waves}
\begin{figure*}
 \centering
 \includegraphics[width=0.99\textwidth]{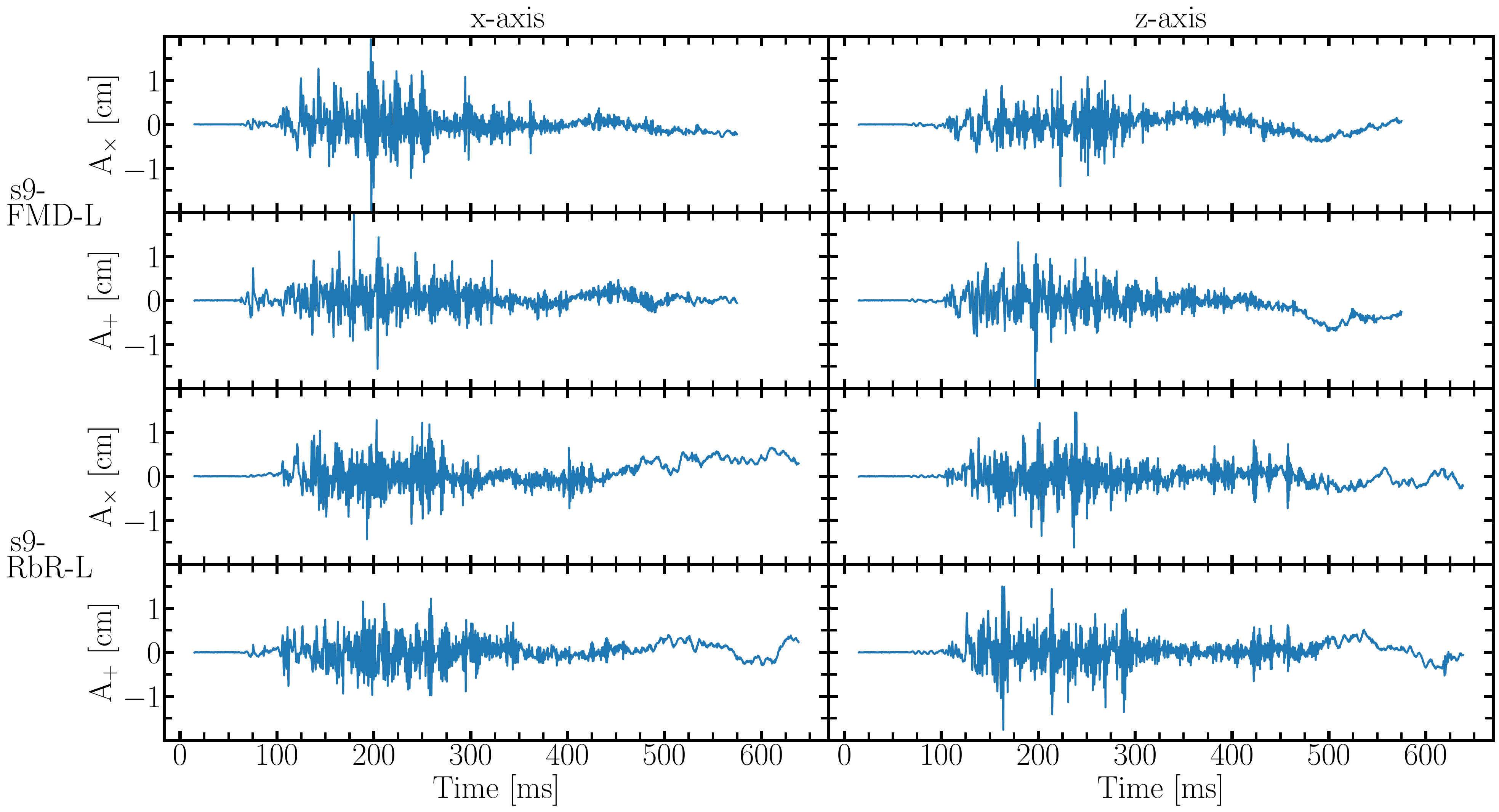}
\caption{GW amplitudes $A_+$ and $A_\times$ as functions of time after core bounce for models
 s9-FMD-L (top two rows) and s9-RbR-L (bottom two rows). The two columns show the amplitudes for
 two observers, one along the z-axis (right) and one along the x-axis (left).}
\label{fig:s9amps_l}
\end{figure*}

In \fig{fig:s9amps_l} we show the GW amplitudes of s9-FMD-L and s9-RbR-L, the corresponding
spectrograms are shown in \fig{fig:s9specs_l}. The amplitudes and spectrograms of s20-FMD-L and
s20-RbR-L are shown in \fig{fig:s20amps_l} and \fig{fig:s20specs_l}, respectively. In the
spectrograms, we over-plot the curves shown in \fig{fig:FMT-RbR}, which helps us compare the typical
characteristics of the signals from the models with high and low resolution. To improve the
readability of the plot, we applied a running Hanning window, 25 ms broad, to smooth the curves from
\fig{fig:FMT-RbR}.

\citet{glas_19} found very similar evolution for all four simulations based on the less massive
progenitor. The average shock radius, the neutrino emission, and the properties of the PNS show a
high degree of similarity in every realisation of the post-bounce evolution of the s9
progenitor. Consequently, the GW signals are also very similar. As in the case of the
high-resolution simulations, the average properties of the emission from models s9-FMD-L and
s9-RbR-L are virtually indistinguishable. We
can see from \fig{fig:spectrum} that the emission in the low-resolution simulations of s9 is
weaker than in the high-resolution runs.

The emission of models s20-FMD-L and s20-RbR-L is similar to their better-resolved counterparts, but
the high-frequency emission almost vanishes
during four distinct periods and is significantly reduced during two periods in model s20-FMD-L. The
four periods of no emission occur at approximately 200, 350, 425, and 575 ms post bounce. The first
epoch of reduced emission starts roughly 500 ms after bounce and the second at 625 ms post
bounce. The signals from model s20-RbR-L only show one time period, around 375 ms post bounce, where
the high-frequency emission completely vanishes. Additionally, the RbR+ model undergoes three
periods of reduced high-frequency emission, at approximately 200, 525 and 600 ms post
bounce. However, the reduction of the intensity in the spectrograms is an order of magnitude weaker
compared to what we see during equivalent episodes in model s20-FMD-L.
As it was the case for the high-resolution models, this emission at low-frequencies is
intermittent because of the alternating periods of SASI and convection in the postshock layer.
The early-time (prior to 250 ms post bounce) emission we saw in the models s20-FMD-H and s20-RbR-H is
also present in the low-resolution models.
The early-time low-frequency emission sets in at 125 ms after bounce in model s20-RbR-L,
approximately 25 ms later than in model
s20-RbR-H. This causes the peak we observed near 100 Hz in
model s20-RbR-H to be shifted down towards 75 Hz in model
s20-RbR-L (see the right top panel of \fig{fig:spectrum_low}).
The relatively strong emission above 100 Hz seen in model s20-RbR-H between 100 ms and 125 ms post bounce is not present in
model s20-RbR-L. When emission starts 25 ms later in s20-RbR-L, the PNS radius has slightly decreased,
which decreases the SASI frequency and shifts the emission peak at 100 Hz
towards 75 Hz. Additionally,
the shock radius in s20-RbR-L is on average larger than the shock
radius in model s20-RbR-H, which further reduces the {\it average}
frequency of the SASI generated emission.

\subsection{Differences induced by decreasing the resolution} \label{sec:resdiff}
\begin{figure*}
 \centering
 \includegraphics[width=0.99\textwidth]{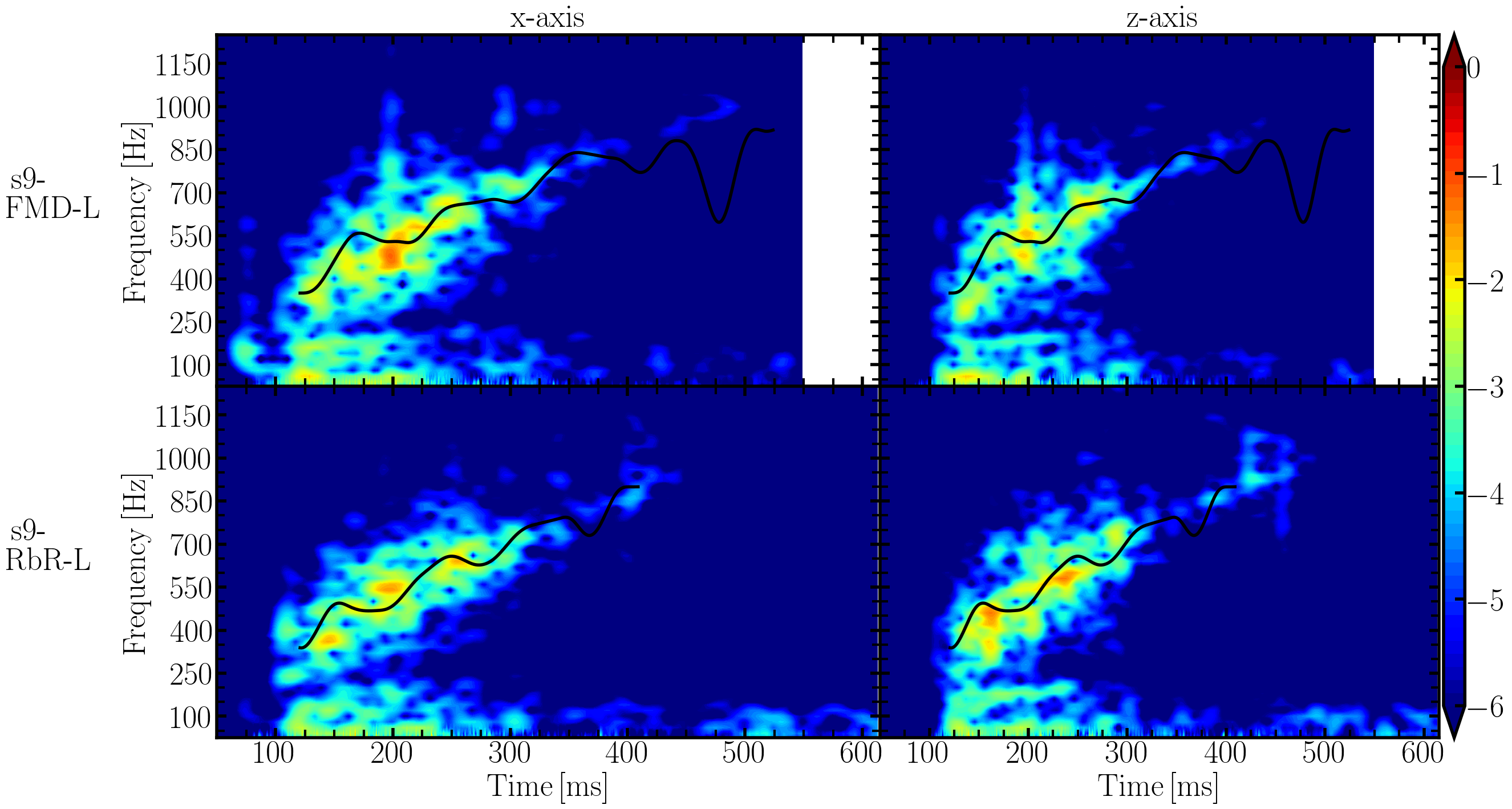}
\caption{Spectrograms, as defined by \eq{eq:spectrogram}, for models s9-FMD-L (top) and s9-RbR-L
 (bottom). Time is given in ms after core bounce. The two columns represent two observers, one
 situated along the z-axis (right) and one along the x-axis (left). The black curves are smoothed
 versions of the curves shown in the right panel of \fig{fig:FMT-RbR}. The colour scale is
 logarithmic. The rate at which data were saved to disk in the simulations means that we have a Nyquist frequency of 1000
Hz, which results in some aliasing (see the last paragraph of section \ref{sec:disscon}).} 
\label{fig:s9specs_l}
\end{figure*}

\begin{figure*}
 \centering
 \includegraphics[width=0.99\textwidth]{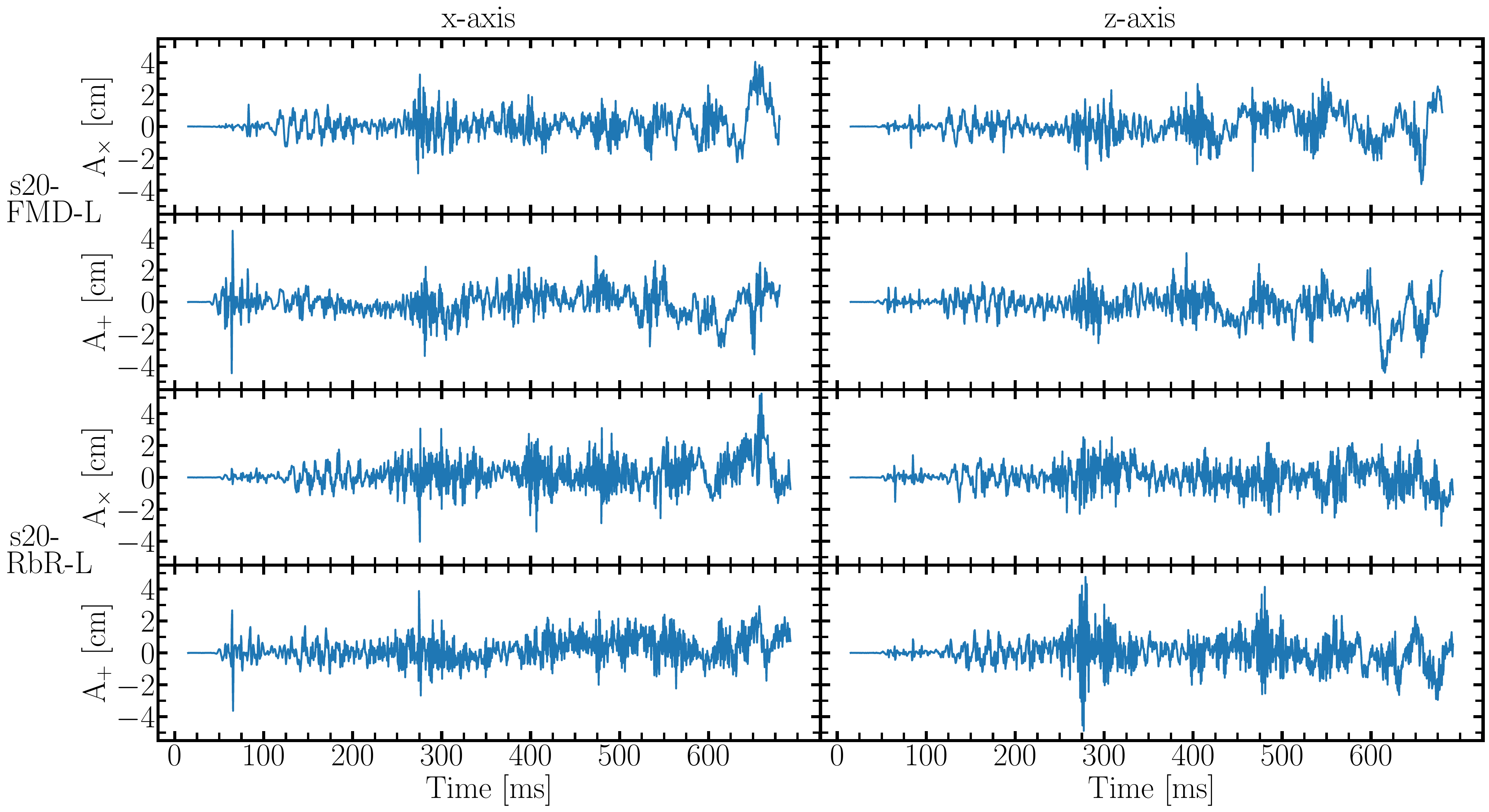} \\
\caption{GW amplitudes $A_+$ and $A_\times$ as functions of time after core bounce for models
  s20-FMD-L (top two rows) and s20-RbR-L (bottom two rows). The two columns show the amplitudes for
 two observers, one along the z-axis (right) and one along the x-axis (left).}
\label{fig:s20amps_l}
\end{figure*}

\begin{figure*}
 \centering
 \includegraphics[width=0.99\textwidth]{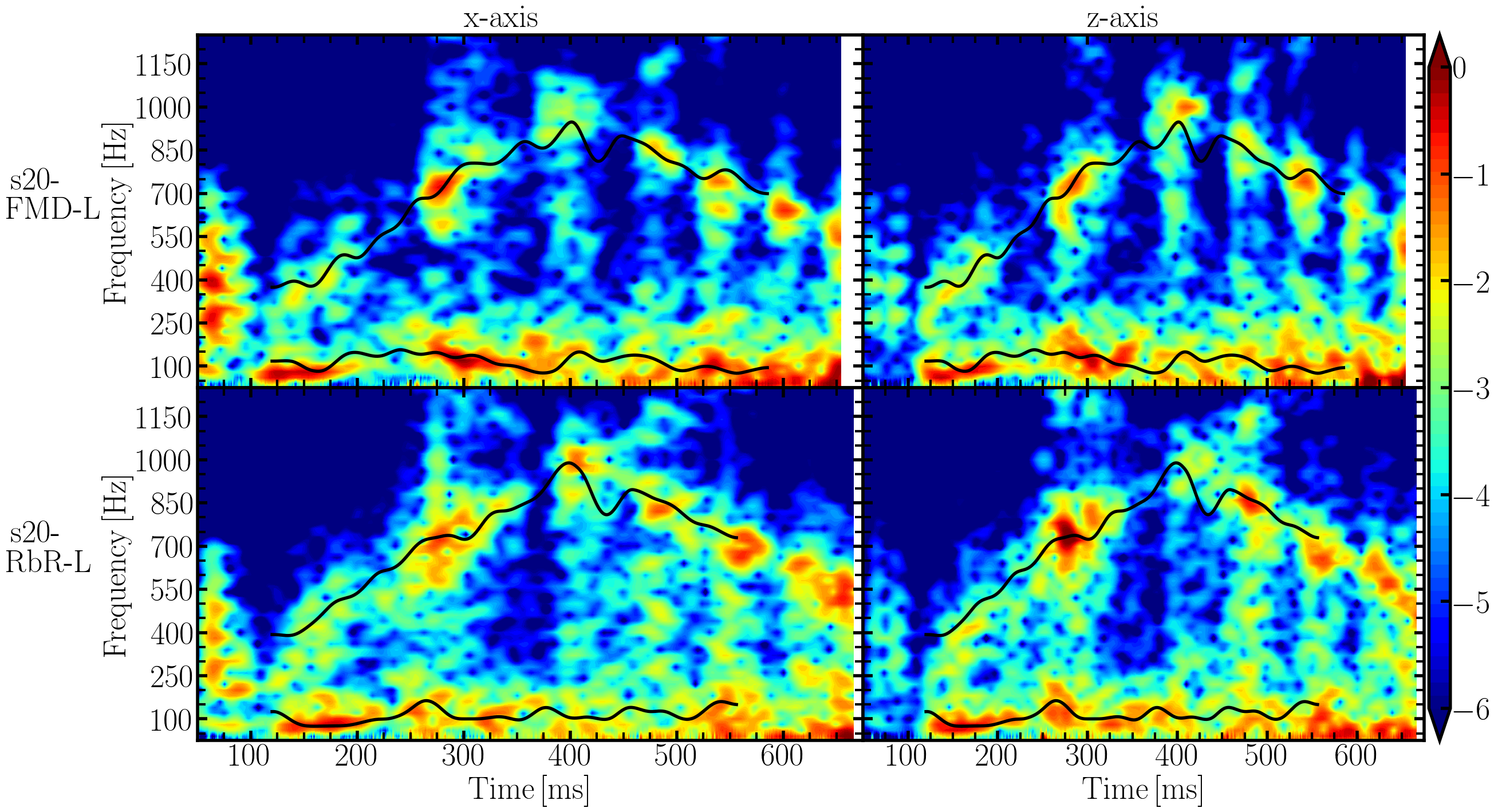}
\caption{Spectrograms, as defined by \eq{eq:spectrogram}, for models s20-FMD-L (top) and s20-RbR-L
 (bottom). Time is given in ms after core bounce. The two columns show the amplitudes for
 two observers, one along the z-axis (right) and one along the x-axis (left). The black curves
 are smoothed versions of the curves shown in the left panel of \fig{fig:FMT-RbR}. The colour scale
 is logarithmic. The rate at which data were saved to disk in the simulations means that we have a Nyquist frequency of 1000
Hz, which results in some aliasing (see the last paragraph of section \ref{sec:disscon}).}
\label{fig:s20specs_l}
\end{figure*}

The peak-frequencies of the GW emission are not drastically affected by changing the resolution.
The overlap of the black curves, which represent the peak-frequencies of the emission from the
corresponding high-resolution simulations, with emission regions in \fig{fig:s9specs_l} and
\fig{fig:s20specs_l} demonstrates that the spectral properties of the high- and low-resolution
models are in good qualitative agreement. \cite{glas_19} did not find significant changes in the
PNS properties between models with different resolution. The shock trajectory of the high-resolution
and low-resolution models show some variability, but the differences in the average shock are only
$\sim\,10$ per cent. These differences are significant in terms of the model dynamics, but not large
enough to have a drastic impact on the typical frequency of the GW emission.

On the other hand, the periods during which the high-frequency signal almost entirely subsides in
model s20-FMD-L point to substantial resolution-dependent differences in the GW signals from
the s20 models. The quiescent periods around 400 and 475 ms post bounce in model s20-FMD-L can be
identified in model s20-FMD-H, too, but the reduction in the GW emission is not as strong. This is
particularly true for the episode starting at 400 ms post bounce.
The phases of weak or suppressed high-frequency emission that we see in the s20 models
are correlated with periods of shock expansion.
Downflows to the PNS become weaker during phases of shock expansion and
thus trigger less high-frequency GW emission from the PNS surface layers.
This suppression of high-frequency emission is weaker in the high-resolution models and
less prominent in the RbR+ simulations.
This is likely due to the stronger convective activity in models with either
higher resolution, which have less numerical viscosity, or with RbR+ transport,
where local hot spots of neutrino heating are stronger drivers of convection in phases when the shock expands.
Both RbR+ and higher resolution therefore facilitate a more continuous emission of high-frequency GWs.

\citet{melson_19} found that the
higher numerical viscosity in low-resolution simulations suppressed the growth of convective
activity. When the shock front contracts and neutrino-driven convection is already disfavored, the
increased numerical viscosity can push the models into a regime where neutrino-driven convection is
significantly reduced in strength
\footnote{Note that we find the same resolution dependence of the
  turbulent energy spectra as \cite{melson_19}.
  Compared to the high-resolution models, the low-resolution models show a
reduction of kinetic energy at small scales ($\ell \ge 10$)
and an increase of energy at scales below $\ell = 10$. We refer the reader to \cite{melson_19} or \cite{nagakura_19} for a
detailed discussion about how resolution affects the turbulent energy spectra and the total kinetic energy in the postshock
layer. In short lowering the resolution decreases the efficiency at which energy
deposited by neutrinos is converted into non-radial kinetic energy.}.
This is reflected in the so-called
$\chi$-parameter. \citet{foglizzo_06} showed that, if the growth of the perturbations is linear,
convection in the postshock layer is disfavored when the $\chi$-parameter is less than three.
Comparing the second to last panel in Fig.~6 of \citet{glas_19} to \fig{fig:s20specs_l}, we see a
clear correlation between low $\chi$ values and quiescent phases in the GW emission above 300 Hz.
In model s20-FMD-L the $\chi$-parameter (see Fig.~6 of \citet{glas_19}) shows prominent minima at
approximately 200 ms and 350 ms post bounce. Both instances are associated with reduced
high-frequency GW emission, with the latter correlating to a period of no high-frequency
radiation. 
The systematically larger average shock radius, which is typically $\sim 5$-$10$ km larger
in model s20-RbR-L than in s20-FMD-L, in combination with large local variations in the neutrino-heating rate
is most likely responsible for keeping convection active in the RbR+ model during periods where emission subsides in model s20-FMD-L.
The conclusion that the RbR+ scheme is at least partly responsible for reducing
the prominence of the quiescent phases is supported by the fact that periods of
weak emission are more pronounced in model s20-FMD-H than in model s20-RbR-H since the shock
trajectories of these two models are very similar.

It is interesting to note that the absence of all but one quiescent phase at 350 ms after bounce in model s20-RbR-L
means that its emission more closely resembles the signals from models s20-FMD-H and s20-RbR-H than s20-FMD-L.

The large average shock radius between 250 and 350 ms post bounce in model s20-RbR-L
lowers the typical frequency associated with the SASI. On the other hand,
it is important to note that the emission is typically spread out in a frequency band of approximately 100 Hz and
it is, therefore, difficult to disentangle short-term shifts in the central frequency of the emission
from inherent stochastic variation.


\section{Summary and Conclusion} \label{sec:disscon}
The main goal of this work was to study the theoretical GW emission from 3D core-collapse supernova
simulations with two different neutrino transport schemes, the RbR+ method and the FMD transport
scheme, to determine if the approximations of the RbR+ method is detrimental for making useful GW
signal predictions, see \cite{skinner_16,just_18} and \cite{glas_19} for discussions about the
potential problems of the RbR+ method. We studied a set of eight models, based on two progenitors.
For each progenitor, the post-bounce evolution was simulated with the two different neutrino transport
methods and at two different resolutions \citep{glas_19}. Consequently, we studied the impact of numerical resolution
and neutrino transport on the GW emission.
Our main findings can be summarised as follows:
\begin{enumerate}
\item{ Differences between the RbR+ and FMD models are, overall, relatively minor for the best-resolved models.
 The spectrograms (Figs. \ref{fig:s9specs} and \ref{fig:s20specs}) display the same basic features.
 Low-frequency GW emission (below $\sim$250-300 Hz) is connected to SASI activity.
 Model s20-RbR-H, and to a lesser degree model s20-FMD-H, show strong low-frequency emission before
 200 ms post bounce. The strong emission is most likely due to a qudrapolar SASI mode which is superimposed
 on the convective flow during the first 200 ms post bounce in the s20 models.
 High-frequency emission originates from g-modes in the layers near the surface of the PNS
 and is mainly driven by accretion downflows.
  The characteristic frequencies of the low- and high-frequency emission agree well
 between all models (\fig{fig:FMT-RbR}).} 
\item{For the s20 models, the RbR+ simulations are more energetic than the FMD simulations and
 the RbR+ simulations emit more GWs at frequencies between the two main signal components
 (Figs.~\ref{fig:s20specs}, \ref{fig:s20specs_l}, and \ref{fig:spectrum}).
 For the s9 models, s9-FMD-H is more energetic than s9-RbR-H (\fig{fig:energyt}).
 Models s9-FMD-L and s9-RbR-L are similar both in terms of their
 spectral energy density and the total emitted energy, with the RbR+ model being
 slightly more energetic (\fig{fig:energyt}).}
\item{Lower resolution, associated with higher numerical viscosity, damps convective mass motions in
 the postshock layer and, consequently, g-mode activity in the PNS surface layer. In s20-FMD-L this leads to the
 appearance of quasi-periodic intermittent phases of GW emission.}
\item{The quasi-periodic suppression of the high-frequency emission seen in the s20 models
  is much weaker in the high-resolution models than in the low-resolution models. We attribute this to the lower numerical
  viscosity in the high-resolution simulations. Furthermore, the suppression is weaker in the RbR+ models,
  which is most likely due to the stronger local variations of the neutrino-heating rate in the RbR+ simulations.
  The local increase in the  heating rate, seen in the RbR+ simulations compared to the FMD models
  \citep{sumiyoshi_15,skinner_16,just_18,glas_19}, provides additional forcing of
 postshock convection and counteracts the suppression of high-frequency emission.}
\end{enumerate}

The main signal features of the GWs from the RbR+ and FMD models are in very good agreement with each other
and it is hard to distinguish between signals from simulations using different neutrino
transport schemes. The amplitude and typical emission frequencies of the two main
signal components agree well in all models. The excellent agreement
is due to the fact that the PNS mass and radius, average shock radius, and preshock
mass accretion rate are almost independent of the neutrino transport method. The average shock radius differs
the most between different models, but the differences are not large enough to
cause any significant shift in the spectral properties of the GW emission. There are differences
in the fine structure of the GW signals, which might be partly stochastic and could also, to a lesser
degree, be linked to the stronger local variations of the neutrino heating in the RbR models.

We did not find any systematic trends as a function of neutrino transport scheme. For example,
based on the s20 simulations one would conclude that the RbR+ simulations were more energetic than the
FMD simulations, but this was not the case for the s9 simulations. A more extensive model set is necessary to uncover
systematic differences in the GW emission from FMD and RbR+ simulations.
Furthermore, the artefacts of the RbR+ simulations decrease with higher resolution, because the finer spacing of the computational
grid causes the local variations in the RbR+ models to shrink spatially.
Overall, changing the resolution has a more significant impact on the model evolution and
differences in the GWs than changing the transport scheme. 

The small differences that we {\it tentatively} concluded to be associated with the RbR+
transport, for example enhanced driving of postshock instabilities by larger variations in the local
neutrino-heating rate compared to FMD models, are likely to become irrelevant once pre-collapse
perturbations in the convectively active burning layers of the progenitor star are included.
In this respect, our study was idealised because the hydrodynamic
instabilities were seeded by small random perturbations (on the level of 0.1\% of the
density in the infall region). These small perturbations in the matter falling through
the stalled shock wave were dominated by the RbR+ heating variations in the postshock layer, whereas
with realistic progenitor perturbations this situation would be reversed.

The signals from core-collapse simulations are inherently stochastic and noisy;
relatively large differences can be seen in predictions made by different groups.
If we aimed to build a catalogue to be used in
matched-filtered searches for laser interferometers, then the differences between the signals
presented in this work would be unacceptable. However, the numerical
complexity and physical uncertainties of the core-collapse problem make such efforts impossible.
A more realistic goal is to produce GW signals that reasonably well predict the essence of emission
from core-collapse supernovae, but with the understanding that the exact details can not be
accurately predicted. The excellent agreement of the central
frequencies (\fig{fig:FMT-RbR}) and typical amplitudes (\fig{fig:s9amps} and \fig{fig:s20amps})
leads us to the conclusion that using the RbR+ approximation is not detrimental to the prediction of
core-collapse supernova GWs.

Finally a note on aliasing. Unfortunately, the sampling rate of our signals was the same as in
\citet{andresen_17} and \citet{andresen_19}. This means that we have a Nyquist frequency of 1000
Hz. Once the Brunt-V\"ais\"ala frequency surpasses this value we lose confidence in the
accuracy of the signals. \cite{radice_19} found that PNS oscillations were the most robust source
of GW emission in their models. \citet{andresen_17} and \citet{andresen_19} concluded that the SASI leads
to strong GW emission, whereas \cite{radice_19} hypothesize that this conclusion is, at least partly, due to
aliasing. It is not clear that the connection can made so readily. The strength of the SASI
generated emission depends strongly on the vigour of the SASI oscillations, the properties of the
flow during the SASI period, and the SASI mode. If strong SASI does not develop, which is the
case for all but one of the models presented by \cite{radice_19}, then naturally the associated GW
signal will not dominate the emission. Ultimately this is a discussion about the development and
nature of the SASI, not one of aliasing.


\section{Acknowledgements}
This work was supported by Deutsche Forschungsgemeinschaft (DFG, German Research Foundation) through
Sonderforschungsbereich (Collaborative Research Centre) SFB-1258 ``Neutrinos and Dark Matter in
Astro- and Particle Physics (NDM)'' and under Germany's Excellence Strategy through Cluster of
Excellence ORIGINS (EXC-2094)---390783311 and by the European Research Council through ERC-AdG
No.\ 341157-COCO2CASA. Computer resources for this project have been provided by the Leibniz
Supercomputing Centre (LRZ) under grant pr62za, and by the Max Planck Computing and Data Facility
(MPCDF) on the HPC system Hydra.

\section{Data Availability}
The GW signals presented in this article
and the underlying quadrupole moments needed to generate signals for arbitrary observer positions
are available in The Garching Core-Collapse Supernova Archive
at \url{https://wwwmpa.mpa-garching.mpg.de/ccsnarchive/archive.html}.

\bibliography{p2ewm}

\begin{thebibliography}{}
\makeatletter
\relax
\def\mn@urlcharsother{\let\do\@makeother \do\$\do\&\do\#\do\^\do\_\do\%\do\~}
\def\mn@doi{\begingroup\mn@urlcharsother \@ifnextchar [ {\mn@doi@}
  {\mn@doi@[]}}
\def\mn@doi@[#1]#2{\def\@tempa{#1}\ifx\@tempa\@empty \href
  {http://dx.doi.org/#2} {doi:#2}\else \href {http://dx.doi.org/#2} {#1}\fi
  \endgroup}
\def\mn@eprint#1#2{\mn@eprint@#1:#2::\@nil}
\def\mn@eprint@arXiv#1{\href {http://arxiv.org/abs/#1} {{\tt arXiv:#1}}}
\def\mn@eprint@dblp#1{\href {http://dblp.uni-trier.de/rec/bibtex/#1.xml}
  {dblp:#1}}
\def\mn@eprint@#1:#2:#3:#4\@nil{\def\@tempa {#1}\def\@tempb {#2}\def\@tempc
  {#3}\ifx \@tempc \@empty \let \@tempc \@tempb \let \@tempb \@tempa \fi \ifx
  \@tempb \@empty \def\@tempb {arXiv}\fi \@ifundefined
  {mn@eprint@\@tempb}{\@tempb:\@tempc}{\expandafter \expandafter \csname
  mn@eprint@\@tempb\endcsname \expandafter{\@tempc}}}

\bibitem[\protect\citeauthoryear{{Abbott} et~al.,}{{Abbott}
  et~al.}{2016}]{ligo_sn_search}
{Abbott} B.~P.,  et~al., 2016, \mn@doi [\prd] {10.1103/PhysRevD.94.102001},
  \href {https://ui.adsabs.harvard.edu/abs/2016PhRvD..94j2001A} {94, 102001}

\bibitem[\protect\citeauthoryear{{Abbott} et~al.,}{{Abbott}
  et~al.}{2020}]{ligo_sn_search2}
{Abbott} B.~P.,  et~al., 2020, \mn@doi [\prd] {10.1103/PhysRevD.101.084002},
  \href {https://ui.adsabs.harvard.edu/abs/2020PhRvD.101h4002A} {101, 084002}

\bibitem[\protect\citeauthoryear{{Andresen}, {M{\"u}ller}, {M{\"u}ller}  \&
  {Janka}}{{Andresen} et~al.}{2017}]{andresen_17}
{Andresen} H.,  {M{\"u}ller} B.,  {M{\"u}ller} E.,   {Janka} H.-T.,  2017,
  \mn@doi [\mnras] {10.1093/mnras/stx618}, \href
  {http://adsabs.harvard.edu/abs/2017MNRAS.468.2032A} {468, 2032}

\bibitem[\protect\citeauthoryear{{Andresen}, {M{\"u}ller}, {Janka}, {Summa},
  {Gill}  \& {Zanolin}}{{Andresen} et~al.}{2019}]{andresen_19}
{Andresen} H.,  {M{\"u}ller} E.,  {Janka} H.~T.,  {Summa} A.,  {Gill} K.,
  {Zanolin} M.,  2019, \mn@doi [Monthly Notices of the Royal Astronomical
  Society] {10.1093/mnras/stz990}, \href
  {https://ui.adsabs.harvard.edu/abs/2019MNRAS.486.2238A} {486, 2238}

\bibitem[\protect\citeauthoryear{{Astone}, {Cerd{\'a}-Dur{\'a}n}, {Di Palma},
  {Drago}, {Muciaccia}, {Palomba}  \& {Ricci}}{{Astone}
  et~al.}{2018}]{astone_18}
{Astone} P.,  {Cerd{\'a}-Dur{\'a}n} P.,  {Di Palma} I.,  {Drago} M.,
  {Muciaccia} F.,  {Palomba} C.,   {Ricci} F.,  2018, \mn@doi [\prd]
  {10.1103/PhysRevD.98.122002}, \href
  {https://ui.adsabs.harvard.edu/abs/2018PhRvD..98l2002A} {98, 122002}

\bibitem[\protect\citeauthoryear{{Blanchet}, {Damour}  \&
  {Sch\"afer}}{{Blanchet} et~al.}{1990}]{blanchet_90}
{Blanchet} L.,  {Damour} T.,   {Sch\"afer} G.,  1990, \mnras, \href
  {http://adsabs.harvard.edu/abs/1990MNRAS.242..289B} {242, 289}

\bibitem[\protect\citeauthoryear{{Blondin} \& {Mezzacappa}}{{Blondin} \&
  {Mezzacappa}}{2006}]{blondin_06}
{Blondin} J.~M.,  {Mezzacappa} A.,  2006, \mn@doi [\apj] {10.1086/500817},
  \href {http://adsabs.harvard.edu/abs/2006ApJ...642..401B} {642, 401}

\bibitem[\protect\citeauthoryear{{Blondin}, {Mezzacappa}  \&
  {DeMarino}}{{Blondin} et~al.}{2003}]{blondin_03}
{Blondin} J.~M.,  {Mezzacappa} A.,   {DeMarino} C.,  2003, \mn@doi [\apj]
  {10.1086/345812}, \href {http://adsabs.harvard.edu/abs/2003ApJ...584..971B}
  {584, 971}

\bibitem[\protect\citeauthoryear{{Bruenn} et~al.,}{{Bruenn}
  et~al.}{2013}]{bruenn_13}
{Bruenn} S.~W.,  et~al., 2013, \mn@doi [\apjl] {10.1088/2041-8205/767/1/L6},
  \href {http://adsabs.harvard.edu/abs/2013ApJ...767L...6B} {767, L6}

\bibitem[\protect\citeauthoryear{{Buras}, {Janka}, {Rampp}  \&
  {Kifonidis}}{{Buras} et~al.}{2006}]{buras_06b}
{Buras} R.,  {Janka} H.-T.,  {Rampp} M.,   {Kifonidis} K.,  2006, \mn@doi
  [\aap] {10.1051/0004-6361:20054654}, \href
  {http://adsabs.harvard.edu/abs/2006A26A...457..281B} {457, 281}

\bibitem[\protect\citeauthoryear{{Finn}}{{Finn}}{1989}]{finn_89}
{Finn} L.~S.,  1989, in {Evans} C.~R.,  {Finn} L.~S.,   {Hobill} D.~W.,  eds,
  Frontiers in Numerical Relativity. Cambridge University Press, Cambridge
  (UK), pp 126--145

\bibitem[\protect\citeauthoryear{{Foglizzo}, {Scheck}  \& {Janka}}{{Foglizzo}
  et~al.}{2006}]{foglizzo_06}
{Foglizzo} T.,  {Scheck} L.,   {Janka} H.-T.,  2006, \mn@doi [\apj]
  {10.1086/508443}, \href {http://adsabs.harvard.edu/abs/2006ApJ...652.1436F}
  {652, 1436}

\bibitem[\protect\citeauthoryear{{Foglizzo}, {Galletti}, {Scheck}  \&
  {Janka}}{{Foglizzo} et~al.}{2007}]{foglizzo_07}
{Foglizzo} T.,  {Galletti} P.,  {Scheck} L.,   {Janka} H.-T.,  2007, \mn@doi
  [\apj] {10.1086/509612}, \href
  {http://adsabs.harvard.edu/abs/2007ApJ...654.1006F} {654, 1006}

\bibitem[\protect\citeauthoryear{{Foglizzo} et~al.,}{{Foglizzo}
  et~al.}{2015}]{foglizzo_15}
{Foglizzo} T.,  et~al., 2015, \mn@doi [\pasa] {10.1017/pasa.2015.9}, \href
  {http://adsabs.harvard.edu/abs/2015PASA...32....9F} {32, 9}

\bibitem[\protect\citeauthoryear{{Gill}, {Wang}, {Valdez}, {Szczepanczyk},
  {Zanolin}  \& {Mukherjee}}{{Gill} et~al.}{2018}]{gill_18}
{Gill} K.,  {Wang} W.,  {Valdez} O.,  {Szczepanczyk} M.,  {Zanolin} M.,
  {Mukherjee} S.,  2018, arXiv e-prints, \href
  {https://ui.adsabs.harvard.edu/abs/2018arXiv180207255G} {p. arXiv:1802.07255}

\bibitem[\protect\citeauthoryear{{Glas}, {Just}, {Janka}  \&
  {Obergaulinger}}{{Glas} et~al.}{2019}]{glas_19}
{Glas} R.,  {Just} O.,  {Janka} H.~T.,   {Obergaulinger} M.,  2019, \mn@doi
  [\apj] {10.3847/1538-4357/ab0423}, \href
  {https://ui.adsabs.harvard.edu/abs/2019ApJ...873...45G} {873, 45}

\bibitem[\protect\citeauthoryear{Gossan, Sutton, Stuver, Zanolin, Gill  \&
  Ott}{Gossan et~al.}{2016}]{gossan_16}
Gossan S.~E.,  Sutton P.,  Stuver A.,  Zanolin M.,  Gill K.,   Ott C.~D.,
  2016, \mn@doi [Phys. Rev. D] {10.1103/PhysRevD.93.042002}, 93, 042002

\bibitem[\protect\citeauthoryear{{Guilet} \& {Foglizzo}}{{Guilet} \&
  {Foglizzo}}{2012}]{guilet_12}
{Guilet} J.,  {Foglizzo} T.,  2012, \mn@doi [\mnras]
  {10.1111/j.1365-2966.2012.20333.x}, \href
  {http://adsabs.harvard.edu/abs/2012MNRAS.421..546G} {421, 546}

\bibitem[\protect\citeauthoryear{{Hanke}, {M{\"u}ller}, {Wongwathanarat},
  {Marek}  \& {Janka}}{{Hanke} et~al.}{2013}]{hanke_13}
{Hanke} F.,  {M{\"u}ller} B.,  {Wongwathanarat} A.,  {Marek} A.,   {Janka}
  H.-T.,  2013, \mn@doi [\apj] {10.1088/0004-637X/770/1/66}, \href
  {http://adsabs.harvard.edu/abs/2013ApJ...770...66H} {770, 66}

\bibitem[\protect\citeauthoryear{Janka}{Janka}{2017}]{janka_17}
Janka H.-T.,  2017, Neutrino-Driven Explosions.
Springer International Publishing, Cham, pp 1095--1150,
  \mn@doi{10.1007/978-3-319-21846-5_109}, \url
  {https://doi.org/10.1007/978-3-319-21846-5_109}

\bibitem[\protect\citeauthoryear{{Just}, {Obergaulinger}  \& {Janka}}{{Just}
  et~al.}{2015}]{just_15}
{Just} O.,  {Obergaulinger} M.,   {Janka} H.~T.,  2015, \mn@doi [\mnras]
  {10.1093/mnras/stv1892}, \href
  {https://ui.adsabs.harvard.edu/abs/2015MNRAS.453.3386J} {453, 3386}

\bibitem[\protect\citeauthoryear{{Just}, {Bollig}, {Janka}, {Obergaulinger},
  {Glas}  \& {Nagataki}}{{Just} et~al.}{2018}]{just_18}
{Just} O.,  {Bollig} R.,  {Janka} H.~T.,  {Obergaulinger} M.,  {Glas} R.,
  {Nagataki} S.,  2018, \mn@doi [\mnras] {10.1093/mnras/sty2578}, \href
  {https://ui.adsabs.harvard.edu/abs/2018MNRAS.481.4786J} {481, 4786}

\bibitem[\protect\citeauthoryear{{Kuroda}, {Takiwaki}  \& {Kotake}}{{Kuroda}
  et~al.}{2016a}]{kuroda_16b}
{Kuroda} T.,  {Takiwaki} T.,   {Kotake} K.,  2016a, \mn@doi [\apjs]
  {10.3847/0067-0049/222/2/20}, \href
  {https://ui.adsabs.harvard.edu/abs/2016ApJS..222...20K} {222, 20}

\bibitem[\protect\citeauthoryear{{Kuroda}, {Kotake}  \& {Takiwaki}}{{Kuroda}
  et~al.}{2016b}]{kuroda_16}
{Kuroda} T.,  {Kotake} K.,   {Takiwaki} T.,  2016b, \mn@doi [\apjl]
  {10.3847/2041-8205/829/1/L14}, \href
  {http://adsabs.harvard.edu/abs/2016ApJ...829L..14K} {829, L14}

\bibitem[\protect\citeauthoryear{{Kuroda}, {Kotake}, {Hayama}  \&
  {Takiwaki}}{{Kuroda} et~al.}{2017}]{kuroda_17}
{Kuroda} T.,  {Kotake} K.,  {Hayama} K.,   {Takiwaki} T.,  2017, \mn@doi [\apj]
  {10.3847/1538-4357/aa988d}, \href
  {http://adsabs.harvard.edu/abs/2017ApJ...851...62K} {851, 62}

\bibitem[\protect\citeauthoryear{{Lentz} et~al.,}{{Lentz}
  et~al.}{2015}]{lentz_15}
{Lentz} E.~J.,  et~al., 2015, \mn@doi [\apjl] {10.1088/2041-8205/807/2/L31},
  \href {http://adsabs.harvard.edu/abs/2015ApJ...807L..31L} {807, L31}

\bibitem[\protect\citeauthoryear{{Logue}, {Ott}, {Heng}, {Kalmus}  \&
  {Scargill}}{{Logue} et~al.}{2012}]{logue_12}
{Logue} J.,  {Ott} C.~D.,  {Heng} I.~S.,  {Kalmus} P.,   {Scargill} J.~H.~C.,
  2012, \mn@doi [\prd] {10.1103/PhysRevD.86.044023}, \href
  {http://adsabs.harvard.edu/abs/2012PhRvD..86d4023L} {86, 044023}

\bibitem[\protect\citeauthoryear{{Marek}, {Dimmelmeier}, {Janka}, {M{\"u}ller}
  \& {Buras}}{{Marek} et~al.}{2006}]{marek_06}
{Marek} A.,  {Dimmelmeier} H.,  {Janka} H.-T.,  {M{\"u}ller} E.,   {Buras} R.,
  2006, \mn@doi [\aap] {10.1051/0004-6361:20052840}, \href
  {http://adsabs.harvard.edu/abs/2006A26A...445..273M} {445, 273}

\bibitem[\protect\citeauthoryear{{Marek}, {Janka}  \& {M{\"u}ller}}{{Marek}
  et~al.}{2009}]{marek_08}
{Marek} A.,  {Janka} H.,   {M{\"u}ller} E.,  2009, \mn@doi [\aap]
  {10.1051/0004-6361/200810883}, \href
  {http://adsabs.harvard.edu/abs/2009A%26A...496..475M} {496, 475}

\bibitem[\protect\citeauthoryear{{Melson}, {Janka}  \& {Marek}}{{Melson}
  et~al.}{2015}]{melson_15a}
{Melson} T.,  {Janka} H.-T.,   {Marek} A.,  2015, \mn@doi [\apjl]
  {10.1088/2041-8205/801/2/L24}, \href
  {http://adsabs.harvard.edu/abs/2015ApJ...801L..24M} {801, L24}

\bibitem[\protect\citeauthoryear{{Melson}, {Kresse}  \& {Janka}}{{Melson}
  et~al.}{2020}]{melson_19}
{Melson} T.,  {Kresse} D.,   {Janka} H.-T.,  2020, \mn@doi [\apj]
  {10.3847/1538-4357/ab72a7}, \href
  {https://ui.adsabs.harvard.edu/abs/2020ApJ...891...27M} {891, 27}

\bibitem[\protect\citeauthoryear{{Mezzacappa} et~al.,}{{Mezzacappa}
  et~al.}{2020}]{mezzacappa_20}
{Mezzacappa} A.,  et~al., 2020, \mn@doi [\prd] {10.1103/PhysRevD.102.023027},
  \href {https://ui.adsabs.harvard.edu/abs/2020PhRvD.102b3027M} {102, 023027}

\bibitem[\protect\citeauthoryear{{Morozova}, {Radice}, {Burrows}  \&
  {Vartanyan}}{{Morozova} et~al.}{2018}]{morozova_18}
{Morozova} V.,  {Radice} D.,  {Burrows} A.,   {Vartanyan} D.,  2018, \mn@doi
  [\apj] {10.3847/1538-4357/aac5f1}, \href
  {http://adsabs.harvard.edu/abs/2018ApJ...861...10M} {861, 10}

\bibitem[\protect\citeauthoryear{{M{\"u}ller}}{{M{\"u}ller}}{2015}]{mueller_15b}
{M{\"u}ller} B.,  2015, \mn@doi [\mnras] {10.1093/mnras/stv1611}, \href
  {http://adsabs.harvard.edu/abs/2015MNRAS.453..287M} {453, 287}

\bibitem[\protect\citeauthoryear{{M{\"u}ller} \& {Janka}}{{M{\"u}ller} \&
  {Janka}}{2014}]{mueller_14}
{M{\"u}ller} B.,  {Janka} H.-T.,  2014, \mn@doi [\apj]
  {10.1088/0004-637X/788/1/82}, \href
  {http://adsabs.harvard.edu/abs/2014ApJ...788...82M} {788, 82}

\bibitem[\protect\citeauthoryear{{M{\"u}ller}, {Janka}  \&
  {Wongwathanarat}}{{M{\"u}ller} et~al.}{2012}]{mueller_e_12}
{M{\"u}ller} E.,  {Janka} H.-T.,   {Wongwathanarat} A.,  2012, \mn@doi [\aap]
  {10.1051/0004-6361/201117611}, \href
  {http://adsabs.harvard.edu/abs/2012A%26A...537A..63M} {537, A63}

\bibitem[\protect\citeauthoryear{{M{\"u}ller}, {Janka}  \&
  {Marek}}{{M{\"u}ller} et~al.}{2013}]{mueller_13}
{M{\"u}ller} B.,  {Janka} H.-T.,   {Marek} A.,  2013, \mn@doi [\apj]
  {10.1088/0004-637X/766/1/43}, \href
  {http://adsabs.harvard.edu/abs/2013ApJ...766...43M} {766, 43}

\bibitem[\protect\citeauthoryear{{M{\"u}ller}, {Melson}, {Heger}  \&
  {Janka}}{{M{\"u}ller} et~al.}{2017}]{mueller_17}
{M{\"u}ller} B.,  {Melson} T.,  {Heger} A.,   {Janka} H.-T.,  2017, \mn@doi
  [\mnras] {10.1093/mnras/stx1962}, \href
  {https://ui.adsabs.harvard.edu/#abs/2017MNRAS.472..491M} {472, 491}

\bibitem[\protect\citeauthoryear{{Murphy}, {Ott}  \& {Burrows}}{{Murphy}
  et~al.}{2009}]{murphy_09}
{Murphy} J.~W.,  {Ott} C.~D.,   {Burrows} A.,  2009, \mn@doi [\apj]
  {10.1088/0004-637X/707/2/1173}, \href
  {http://adsabs.harvard.edu/abs/2009ApJ...707.1173M} {707, 1173}

\bibitem[\protect\citeauthoryear{{Nagakura}, {Burrows}, {Radice}  \&
  {Vartanyan}}{{Nagakura} et~al.}{2019}]{nagakura_19}
{Nagakura} H.,  {Burrows} A.,  {Radice} D.,   {Vartanyan} D.,  2019, \mn@doi
  [\mnras] {10.1093/mnras/stz2730}, \href
  {https://ui.adsabs.harvard.edu/abs/2019MNRAS.490.4622N} {490, 4622}

\bibitem[\protect\citeauthoryear{{O'Connor} \& {Couch}}{{O'Connor} \&
  {Couch}}{2018}]{oconnor_18}
{O'Connor} E.~P.,  {Couch} S.~M.,  2018, \mn@doi [The Astrophysical Journal]
  {10.3847/1538-4357/aadcf7}, \href
  {https://ui.adsabs.harvard.edu/abs/2018ApJ...865...81O} {865, 81}

\bibitem[\protect\citeauthoryear{Obergaulinger}{Obergaulinger}{2008}]{obergaulinger_phd}
Obergaulinger M.,  2008, Dissertation, Technische Universität München,
  München

\bibitem[\protect\citeauthoryear{{Ohnishi}, {Kotake}  \& {Yamada}}{{Ohnishi}
  et~al.}{2006}]{ohnishi_06}
{Ohnishi} N.,  {Kotake} K.,   {Yamada} S.,  2006, \mn@doi [\apj]
  {10.1086/500554}, \href {http://adsabs.harvard.edu/abs/2006ApJ...641.1018O}
  {641, 1018}

\bibitem[\protect\citeauthoryear{{Ohnishi}, {Iwakami}, {Kotake}, {Yamada},
  {Fujioka}  \& {Takabe}}{{Ohnishi} et~al.}{2008}]{ohnishi_08}
{Ohnishi} N.,  {Iwakami} W.,  {Kotake} K.,  {Yamada} S.,  {Fujioka} S.,
  {Takabe} H.,  2008, \mn@doi [Journal of Physics Conference Series]
  {10.1088/1742-6596/112/4/042018}, \href
  {http://adsabs.harvard.edu/abs/2008JPhCS.112d2018O} {112, 042018}

\bibitem[\protect\citeauthoryear{Oohara, Nakamura  \& Shibata}{Oohara
  et~al.}{1997}]{oohara_97}
Oohara K.-i.,  Nakamura T.,   Shibata M.,  1997, \mn@doi [Progress of
  Theoretical Physics Supplement] {10.1143/PTPS.128.183}, 128, 183

\bibitem[\protect\citeauthoryear{{Ott}, {Roberts}, {da Silva Schneider},
  {Fedrow}, {Haas}  \& {Schnetter}}{{Ott} et~al.}{2018}]{ott_18}
{Ott} C.~D.,  {Roberts} L.~F.,  {da Silva Schneider} A.,  {Fedrow} J.~M.,
  {Haas} R.,   {Schnetter} E.,  2018, \mn@doi [\apjl]
  {10.3847/2041-8213/aaa967}, \href
  {https://ui.adsabs.harvard.edu/abs/2018ApJ...855L...3O} {855, L3}

\bibitem[\protect\citeauthoryear{{Powell}}{{Powell}}{2018}]{powell_18}
{Powell} J.,  2018, \mn@doi [Classical and Quantum Gravity]
  {10.1088/1361-6382/aacf18}, \href
  {https://ui.adsabs.harvard.edu/abs/2018CQGra..35o5017P} {35, 155017}

\bibitem[\protect\citeauthoryear{{Powell} \& {M{\"u}ller}}{{Powell} \&
  {M{\"u}ller}}{2019}]{powell_19}
{Powell} J.,  {M{\"u}ller} B.,  2019, \mn@doi [\mnras] {10.1093/mnras/stz1304},
  \href {https://ui.adsabs.harvard.edu/abs/2019MNRAS.487.1178P} {487, 1178}

\bibitem[\protect\citeauthoryear{{Powell} \& {M{\"u}ller}}{{Powell} \&
  {M{\"u}ller}}{2020}]{powell_20}
{Powell} J.,  {M{\"u}ller} B.,  2020, \mn@doi [\mnras]
  {10.1093/mnras/staa1048}, \href
  {https://ui.adsabs.harvard.edu/abs/2020MNRAS.tmp.1245P} {}

\bibitem[\protect\citeauthoryear{{Powell}, {Gossan}, {Logue}  \&
  {Heng}}{{Powell} et~al.}{2016}]{powell_16}
{Powell} J.,  {Gossan} S.~E.,  {Logue} J.,   {Heng} I.~S.,  2016, \mn@doi
  [\prd] {10.1103/PhysRevD.94.123012}, \href
  {https://ui.adsabs.harvard.edu/abs/2016PhRvD..94l3012P} {94, 123012}

\bibitem[\protect\citeauthoryear{{Powell}, {Szczepanczyk}  \& {Heng}}{{Powell}
  et~al.}{2017}]{powell_17}
{Powell} J.,  {Szczepanczyk} M.,   {Heng} I.~S.,  2017, \mn@doi [\prd]
  {10.1103/PhysRevD.96.123013}, \href
  {https://ui.adsabs.harvard.edu/#abs/2017PhRvD..96l3013P} {96, 123013}

\bibitem[\protect\citeauthoryear{{Radice}, {Morozova}, {Burrows}, {Vartanyan}
  \& {Nagakura}}{{Radice} et~al.}{2019}]{radice_19}
{Radice} D.,  {Morozova} V.,  {Burrows} A.,  {Vartanyan} D.,   {Nagakura} H.,
  2019, \mn@doi [\apjl] {10.3847/2041-8213/ab191a}, \href
  {https://ui.adsabs.harvard.edu/abs/2019ApJ...876L...9R} {876, L9}

\bibitem[\protect\citeauthoryear{{Rampp} \& {Janka}}{{Rampp} \&
  {Janka}}{2002}]{rampp_02}
{Rampp} M.,  {Janka} H.-T.,  2002, \mn@doi [\aap] {10.1051/0004-6361:20021398},
  \href
  {http://adsabs.harvard.edu/cgi-bin/nph-bib_query?bibcode=2002A26A...396..361R&db_key=AST}
  {396, 361}

\bibitem[\protect\citeauthoryear{{Roberts}, {Ott}, {Haas}, {O'Connor}, {Diener}
   \& {Schnetter}}{{Roberts} et~al.}{2016}]{roberts_16}
{Roberts} L.~F.,  {Ott} C.~D.,  {Haas} R.,  {O'Connor} E.~P.,  {Diener} P.,
  {Schnetter} E.,  2016, \mn@doi [\apj] {10.3847/0004-637X/831/1/98}, \href
  {https://ui.adsabs.harvard.edu/abs/2016ApJ...831...98R} {831, 98}

\bibitem[\protect\citeauthoryear{{Roma}, {Powell}, {Heng}  \& {Frey}}{{Roma}
  et~al.}{2019}]{roma_19}
{Roma} V.,  {Powell} J.,  {Heng} I.~S.,   {Frey} R.,  2019, \mn@doi [\prd]
  {10.1103/PhysRevD.99.063018}, \href
  {https://ui.adsabs.harvard.edu/abs/2019PhRvD..99f3018R} {99, 063018}

\bibitem[\protect\citeauthoryear{{Scheck}, {Janka}, {Foglizzo}  \&
  {Kifonidis}}{{Scheck} et~al.}{2008}]{scheck_08}
{Scheck} L.,  {Janka} H.-T.,  {Foglizzo} T.,   {Kifonidis} K.,  2008, \mn@doi
  [\aap] {10.1051/0004-6361:20077701}, \href
  {http://adsabs.harvard.edu/abs/2008A26A...477..931S} {477, 931}

\bibitem[\protect\citeauthoryear{{Skinner}, {Burrows}  \& {Dolence}}{{Skinner}
  et~al.}{2016}]{skinner_16}
{Skinner} M.~A.,  {Burrows} A.,   {Dolence} J.~C.,  2016, \mn@doi [\apj]
  {10.3847/0004-637X/831/1/81}, \href
  {http://adsabs.harvard.edu/abs/2016ApJ...831...81S} {831, 81}

\bibitem[\protect\citeauthoryear{{Sotani} \& {Takiwaki}}{{Sotani} \&
  {Takiwaki}}{2020}]{sotani_20}
{Sotani} H.,  {Takiwaki} T.,  2020, \mn@doi [\prd]
  {10.1103/PhysRevD.102.023028}, \href
  {https://ui.adsabs.harvard.edu/abs/2020PhRvD.102b3028S} {102, 023028}

\bibitem[\protect\citeauthoryear{{Sotani}, {Kuroda}, {Takiwaki}  \&
  {Kotake}}{{Sotani} et~al.}{2017}]{sotani_17}
{Sotani} H.,  {Kuroda} T.,  {Takiwaki} T.,   {Kotake} K.,  2017, \mn@doi [\prd]
  {10.1103/PhysRevD.96.063005}, \href
  {https://ui.adsabs.harvard.edu/abs/2017PhRvD..96f3005S} {96, 063005}

\bibitem[\protect\citeauthoryear{{Steiner}, {Hempel}  \& {Fischer}}{{Steiner}
  et~al.}{2013}]{steiner_13}
{Steiner} A.~W.,  {Hempel} M.,   {Fischer} T.,  2013, \mn@doi [\apj]
  {10.1088/0004-637X/774/1/17}, \href
  {https://ui.adsabs.harvard.edu/abs/2013ApJ...774...17S} {774, 17}

\bibitem[\protect\citeauthoryear{{Sukhbold}, {Ertl}, {Woosley}, {Brown}  \&
  {Janka}}{{Sukhbold} et~al.}{2016}]{sukhbold_16}
{Sukhbold} T.,  {Ertl} T.,  {Woosley} S.~E.,  {Brown} J.~M.,   {Janka} H.~T.,
  2016, \mn@doi [\apj] {10.3847/0004-637X/821/1/38}, \href
  {https://ui.adsabs.harvard.edu/abs/2016ApJ...821...38S} {821, 38}

\bibitem[\protect\citeauthoryear{{Sumiyoshi}, {Takiwaki}, {Matsufuru}  \&
  {Yamada}}{{Sumiyoshi} et~al.}{2015}]{sumiyoshi_15}
{Sumiyoshi} K.,  {Takiwaki} T.,  {Matsufuru} H.,   {Yamada} S.,  2015, \mn@doi
  [\apjs] {10.1088/0067-0049/216/1/5}, \href
  {https://ui.adsabs.harvard.edu/abs/2015ApJS..216....5S} {216, 5}

\bibitem[\protect\citeauthoryear{{Summa}, {Janka}, {Melson}  \&
  {Marek}}{{Summa} et~al.}{2018}]{summa_18}
{Summa} A.,  {Janka} H.-T.,  {Melson} T.,   {Marek} A.,  2018, \mn@doi [\apj]
  {10.3847/1538-4357/aa9ce8}, \href
  {http://adsabs.harvard.edu/abs/2018ApJ...852...28S} {852, 28}

\bibitem[\protect\citeauthoryear{{Suvorova}, {Powell}  \& {Melatos}}{{Suvorova}
  et~al.}{2019}]{suvorova_19}
{Suvorova} S.,  {Powell} J.,   {Melatos} A.,  2019, \mn@doi [\prd]
  {10.1103/PhysRevD.99.123012}, \href
  {https://ui.adsabs.harvard.edu/abs/2019PhRvD..99l3012S} {99, 123012}

\bibitem[\protect\citeauthoryear{{Takiwaki}, {Kotake}  \& {Suwa}}{{Takiwaki}
  et~al.}{2012}]{takiwaki_12}
{Takiwaki} T.,  {Kotake} K.,   {Suwa} Y.,  2012, \mn@doi [\apj]
  {10.1088/0004-637X/749/2/98}, \href
  {http://adsabs.harvard.edu/abs/2012ApJ...749...98T} {749, 98}

\bibitem[\protect\citeauthoryear{{Takiwaki}, {Kotake}  \& {Suwa}}{{Takiwaki}
  et~al.}{2014}]{takiwaki_14}
{Takiwaki} T.,  {Kotake} K.,   {Suwa} Y.,  2014, \mn@doi [\apj]
  {10.1088/0004-637X/786/2/83}, \href
  {http://adsabs.harvard.edu/abs/2014ApJ...786...83T} {786, 83}

\bibitem[\protect\citeauthoryear{{Torres-Forn{\'e}}, {Cerd{\'a}-Dur{\'a}n},
  {Passamonti}  \& {Font}}{{Torres-Forn{\'e}} et~al.}{2018}]{torres-forne_18}
{Torres-Forn{\'e}} A.,  {Cerd{\'a}-Dur{\'a}n} P.,  {Passamonti} A.,   {Font}
  J.~A.,  2018, \mn@doi [\mnras] {10.1093/mnras/stx3067}, \href
  {https://ui.adsabs.harvard.edu/abs/2018MNRAS.474.5272T} {474, 5272}

\bibitem[\protect\citeauthoryear{{Torres-Forn{\'e}}, {Cerd{\'a}-Dur{\'a}n},
  {Obergaulinger}, {M{\"u}ller}  \& {Font}}{{Torres-Forn{\'e}}
  et~al.}{2019a}]{torres-forne_19a}
{Torres-Forn{\'e}} A.,  {Cerd{\'a}-Dur{\'a}n} P.,  {Obergaulinger} M.,
  {M{\"u}ller} B.,   {Font} J.~A.,  2019a, \mn@doi [\prl]
  {10.1103/PhysRevLett.123.051102}, \href
  {https://ui.adsabs.harvard.edu/abs/2019PhRvL.123e1102T} {123, 051102}

\bibitem[\protect\citeauthoryear{{Torres-Forn{\'e}}, {Cerd{\'a}-Dur{\'a}n},
  {Passamonti}, {Obergaulinger}  \& {Font}}{{Torres-Forn{\'e}}
  et~al.}{2019b}]{torres-forne_19b}
{Torres-Forn{\'e}} A.,  {Cerd{\'a}-Dur{\'a}n} P.,  {Passamonti} A.,
  {Obergaulinger} M.,   {Font} J.~A.,  2019b, \mn@doi [\mnras]
  {10.1093/mnras/sty2854}, \href
  {https://ui.adsabs.harvard.edu/abs/2019MNRAS.482.3967T} {482, 3967}

\bibitem[\protect\citeauthoryear{{Vartanyan}, {Burrows}  \&
  {Radice}}{{Vartanyan} et~al.}{2019}]{vartanyan_19}
{Vartanyan} D.,  {Burrows} A.,   {Radice} D.,  2019, \mn@doi [\mnras]
  {10.1093/mnras/stz2307}, \href
  {https://ui.adsabs.harvard.edu/abs/2019MNRAS.489.2227V} {489, 2227}

\bibitem[\protect\citeauthoryear{{Wongwathanarat}, {Janka}  \&
  {M{\"u}ller}}{{Wongwathanarat} et~al.}{2010}]{wongwathanarat_10b}
{Wongwathanarat} A.,  {Janka} H.,   {M{\"u}ller} E.,  2010, \mn@doi [\apjl]
  {10.1088/2041-8205/725/1/L106}, \href
  {http://adsabs.harvard.edu/abs/2010ApJ...725L.106W} {725, L106}

\bibitem[\protect\citeauthoryear{{Wongwathanarat}, {M{\"u}ller}  \&
  {Janka}}{{Wongwathanarat} et~al.}{2015}]{wongwathanarat_15}
{Wongwathanarat} A.,  {M{\"u}ller} E.,   {Janka} H.-T.,  2015, \mn@doi [\aap]
  {10.1051/0004-6361/201425025}, \href
  {http://adsabs.harvard.edu/abs/2015A%26A...577A..48W} {577, A48}

\bibitem[\protect\citeauthoryear{{Woosley} \& {Heger}}{{Woosley} \&
  {Heger}}{2007}]{woosley_07}
{Woosley} S.~E.,  {Heger} A.,  2007, \mn@doi [\physrep]
  {10.1016/j.physrep.2007.02.009}, \href
  {http://adsabs.harvard.edu/abs/2007PhR...442..269W} {442, 269}

\bibitem[\protect\citeauthoryear{{Woosley} \& {Heger}}{{Woosley} \&
  {Heger}}{2015}]{woosley_15}
{Woosley} S.~E.,  {Heger} A.,  2015, \mn@doi [\apj]
  {10.1088/0004-637X/810/1/34}, \href
  {https://ui.adsabs.harvard.edu/abs/2015ApJ...810...34W} {810, 34}

\bibitem[\protect\citeauthoryear{{Yakunin} et~al.,}{{Yakunin}
  et~al.}{2015}]{yakunin_15}
{Yakunin} K.~N.,  et~al., 2015, \mn@doi [\prd] {10.1103/PhysRevD.92.084040},
  \href {http://adsabs.harvard.edu/abs/2015PhRvD..92h4040Y} {92, 084040}

\bibitem[\protect\citeauthoryear{{van Putten}}{{van
  Putten}}{2016}]{vanputten_16}
{van Putten} M. H.~P.~M.,  2016, \mn@doi [\apj] {10.3847/0004-637X/819/2/169},
  \href {https://ui.adsabs.harvard.edu/abs/2016ApJ...819..169V} {819, 169}

\makeatother
\end{thebibliography}

\end{document}